\documentclass[twoside,twocolumn,9pt]{article}
\usepackage{extsizes}
\usepackage[super,sort&compress,comma]{natbib} 
\usepackage[version=3]{mhchem}
\usepackage[left=1.5cm, right=1.5cm, top=1.785cm, bottom=2.0cm]{geometry}
\usepackage{balance}
\usepackage{mathptmx}
\usepackage{sectsty}
\usepackage{graphicx} 
\usepackage{lastpage}
\usepackage[format=plain,justification=justified,singlelinecheck=false,font={stretch=1.125,small,sf},labelfont=bf,labelsep=space]{caption}
\usepackage{float}
\usepackage{fancyhdr}
\usepackage{fnpos}
\usepackage[english]{babel}
\addto{\captionsenglish}{%
  
}
\usepackage{array}
\usepackage{droidsans}
\usepackage{charter}
\usepackage[T1]{fontenc}
\usepackage[usenames,dvipsnames]{xcolor}
\usepackage{setspace}
\usepackage[compact]{titlesec}
\usepackage{hyperref}
\usepackage{acronym} % used for adding accronyms
\usepackage{textcomp, gensymb}
\usepackage[version=3]{mhchem}
\usepackage{amssymb}
\usepackage{diagbox}
\usepackage{multirow}
\usepackage{graphicx}
\usepackage{amsmath}
\RequirePackage{color}
\usepackage[utf8]{inputenc}
\usepackage[flushleft]{threeparttable}

\acrodef{LIB}[LIB]{lithium-ion battery}
\newacroplural{LIB}[LIBs]{lithium-ion batteries}
\acrodef{RAZIB}[RAZIB]{rechargeable aqueous zinc-ion battery}
\newacroplural{RAZIB}[RAZIBs]{Rechargeable aqueous zinc-ion batteries}
\acrodef{OER}[OER]{oxygen evolution reaction}
\acrodef{HER}[HER]{hydrogen evolution reaction}
\acrodef{SHE}[SHE]{standard hydrogen electrode}
\acrodef{OCV}[OCV]{open-circuit voltage} 
\acrodef{GSES}[GSES]{grid-scale energy storage}

\acrodef{SI}[SI]{supplementary information}

\acrodef{GCD}[GCD]{galvanostatic charge/discharge}
\acrodef{CV}[CV]{cyclic voltammetry}
\acrodef{JTE}[JTE]{Jahn-Teller effect}
\acrodef{1HNMR}[\ce{^{1}H} NMR]{\ce{^{1}H} nuclear magnetic resonance}
\acrodef{NMR}[NMR]{nuclear magnetic resonance}
\acrodef{SEM}[SEM]{scanning electron microscopy} 
\acrodef{Mn2+conc}[$\text{[}$\ce{Mn^{2+}_{(aq)}}$\text{]}$]{\ce{Mn^{2+}_{(aq)}} concentration} 
\acrodef{PTFE}[PTFE]{polytetrafluoroethylene} 

\acrodef{DFT}[DFT]{density functional theory}
\acrodef{GGA}[GGA]{generalized gradient approximation}
\acrodef{PBE}[PBE]{Perdew–Burke-Ernzerhof}
\acrodef{VBE}[VBE]{valence band edge}
\acrodef{CBE}[CBE]{conduction band edge}
\acrodef{EF}[$E_{\mathrm{F}}$]{Fermi energy}
\acrodef{dEF}[${\mathrm{\Delta}}E_{\mathrm{F}}$]{Fermi energy variation}
\acrodef{VASP}[VASP]{Vienna \textit{ab-initio} simulation package}
\acrodef{SC}[SC]{self-consistent}
\acrodef{FNV}[FNV]{Freysoldt-Neugebauer-Van de Walle}
\acrodef{locpot}[\textit{U}$_{\text{local}}$]{local potential}

\acrodef{VMn}[V$_{\ce{Mn}}$]{\ce{Mn} vacancy}
\acrodef{VMn1}[V$_{\ce{Mn}1}$]{\ce{Mn}1 vacancy}
\acrodef{VMn2}[V$_{\ce{Mn}2}$]{\ce{Mn}2 vacancy}
\acrodef{EdVMn}[\textit{E}$_{d}$(V$_{\ce{Mn}}^{q}$)]{\ce{Mn} vacancy formation energy}
\acrodef{EdVMn0}[\textit{E}$_{d}$(V$_{\ce{Mn}}^{0}$)]{neutral \ce{Mn} vacancy formation energy}
\acrodef{EdVMn-2CBE}[\textit{E}$_{d}$(V$_{\ce{Mn}}^{-2}$,CBE)]{\ce{Mn} vacancy formation energy for a $q=-2$ charged defect at the \acl{CBE}}

\acrodef{dErelax}[$\Delta$\textit{E}$_{\text{rlx}}$(V$_{\ce{Mn}}^{0}$)]{cell relaxation energy variation for neutral \ce{Mn} vacancy}
\acrodef{EdVMn0relax}[\textit{E}$_{d}$(V$_{\ce{Mn}}^{0}$)$_\text{wor}$]{neutral \ce{Mn} vacancy formation energy without relaxation}
\acrodef{VMn0}[V$_{\ce{Mn}}^{0}$]{neutral \ce{Mn} vacancy}

\acrodef{phiSHEMn}[$\phi_{\text{SHE}}$]{electrode potential vs the \ac{SHE}}
\acrodef{phiZnMn}[$\phi_{\text{Zn}}$]{electrode potential vs \ce{Zn}/\ce{Zn^{2+}}}
\acrodef{phiZnMnrev}[$\phi^{\text{rev}}_{\text{Zn}}$(V$^{~0}_{\ce{Mn}}$)]{reversible electrode potential for \ac{VMn0} formation}
\acrodef{phiSHEZnstd}[$\phi_{\mathrm{SHE}}(\ce{Zn}/\ce{Zn^{2+}})$]{standard electrode potential for the \ce{Zn}/\ce{Zn^{2+}} redox couple vs \ac{SHE}}
\acrodef{phi}[$\phi$]{electrode potential}

\newcommand*{\vrectangleA}{{\ooalign{\lower.3ex\hbox{$\sqcup$}\cr\raise.4ex\hbox{$\sqcap$}}}}

\usepackage{epstopdf}%This line makes .eps figures into .pdf - please comment out if not required.

\definecolor{cream}{RGB}{222,217,201}

\begin{document}

\pagestyle{fancy}
\thispagestyle{plain}
\fancypagestyle{plain}{
%%%HEADER%%%
\renewcommand{\headrulewidth}{0pt}
}
%%%END OF HEADER%%%

%%%PAGE SETUP - Please do not change any commands within this section%%%
\makeFNbottom
\makeatletter
\renewcommand\LARGE{\@setfontsize\LARGE{15pt}{17}}
\renewcommand\Large{\@setfontsize\Large{12pt}{14}}
\renewcommand\large{\@setfontsize\large{10pt}{12}}
\renewcommand\footnotesize{\@setfontsize\footnotesize{7pt}{10}}
\makeatother

\renewcommand{\thefootnote}{\fnsymbol{footnote}}
\renewcommand\footnoterule{\vspace*{1pt}% 
\color{cream}\hrule width 3.5in height 0.4pt \color{black}\vspace*{5pt}} 
\setcounter{secnumdepth}{5}

\makeatletter 
\renewcommand\@biblabel[1]{#1}            
\renewcommand\@makefntext[1]% 
{\noindent\makebox[0pt][r]{\@thefnmark\,}#1}
\makeatother 
\renewcommand{\figurename}{\small{Fig.}~}
\sectionfont{\sffamily\Large}
\subsectionfont{\normalsize}
\subsubsectionfont{\bf}
\setstretch{1.125} %In particular, please do not alter this line.
\setlength{\skip\footins}{0.8cm}
\setlength{\footnotesep}{0.25cm}
\setlength{\jot}{10pt}
\titlespacing*{\section}{0pt}{4pt}{4pt}
\titlespacing*{\subsection}{0pt}{15pt}{1pt}
%%%END OF PAGE SETUP%%%

%%%FOOTER%%%
\fancyfoot{}
\fancyfoot[LO,RE]{\vspace{-7.1pt}\includegraphics[height=9pt]{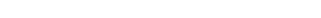}}
\fancyfoot[CO]{\vspace{-7.1pt}\hspace{13.2cm}\includegraphics{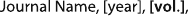}}
\fancyfoot[CE]{\vspace{-7.2pt}\hspace{-14.2cm}\includegraphics{head_foot/RF}}
\fancyfoot[RO]{\footnotesize{\sffamily{1--\pageref{LastPage} ~\textbar  \hspace{2pt}\thepage}}}
\fancyfoot[LE]{\footnotesize{\sffamily{\thepage~\textbar\hspace{3.45cm} 1--\pageref{LastPage}}}}
\fancyhead{}
\renewcommand{\headrulewidth}{0pt} 
\renewcommand{\footrulewidth}{0pt}
\setlength{\arrayrulewidth}{1pt}
\setlength{\columnsep}{6.5mm}
\setlength\bibsep{1pt}
%%%END OF FOOTER%%%

%%%FIGURE SETUP - please do not change any commands within this section%%%
\makeatletter 
\newlength{\figrulesep} 
\setlength{\figrulesep}{0.5\textfloatsep} 

\newcommand{\topfigrule}{\vspace*{-1pt}% 
\noindent{\color{cream}\rule[-\figrulesep]{\columnwidth}{1.5pt}} }

\newcommand{\botfigrule}{\vspace*{-2pt}% 
\noindent{\color{cream}\rule[\figrulesep]{\columnwidth}{1.5pt}} }

\newcommand{\dblfigrule}{\vspace*{-1pt}% 
\noindent{\color{cream}\rule[-\figrulesep]{\textwidth}{1.5pt}} }

\makeatother
%%%END OF FIGURE SETUP%%%

%%%TITLE, AUTHORS AND ABSTRACT%%%
\twocolumn[
  \begin{@twocolumnfalse}
{\includegraphics[height=30pt]{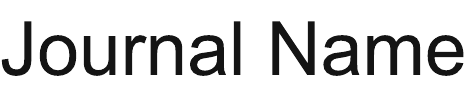}\hfill\raisebox{0pt}[0pt][0pt]{\includegraphics[height=55pt]{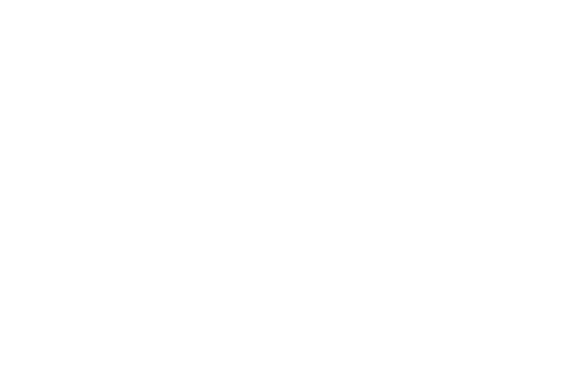}}\\[1ex]
\includegraphics[width=18.5cm]{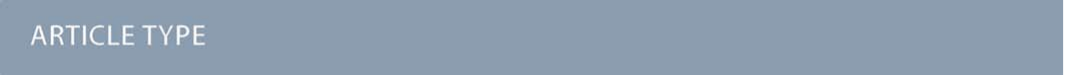}}\par
\vspace{1em}
\sffamily
\begin{tabular}{m{4.5cm} p{13.5cm} }

\includegraphics{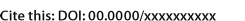} & \noindent\LARGE{\textbf{Unveiling the origin of the capacity fade in \ce{MnO2} zinc-ion battery cathodes through an analysis of the \ce{Mn} vacancy formation\dag}} \\%Article title goes here instead of the text "This is the title"
\vspace{0.3cm} & \vspace{0.3cm} \\

 & \noindent\large{Caio Miranda Miliante,$^{\ast}$\textit{$^{a}$} Kevin J. Sanders,\textit{$^{b}$} Liam J. McGoldrick,\textit{$^{b}$} Nicola Seriani,\textit{$^{c}$} Brian D. Adams,\textit{$^{d}$} Gillian R. Goward,\textit{$^{b}$} Drew Higgins,\textit{$^{e}$} and Oleg Rubel\textit{$^{a}$}} \\%Author names go here instead of "Full name", etc.

\includegraphics{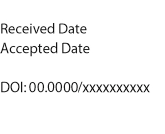} & \noindent\normalsize{Currently explored \ac{RAZIB} cathode materials, such as $\alpha$-\ce{MnO2}, suffer from severe capacity fade when charging and discharging at rates appropriate for grid-scale operation. \ce{Mn} dissolution has been previously identified as the cause of $\alpha$-\ce{MnO2} cathode degradation during \ac{RAZIB} cycling, with conflicting evidence being found in support of the proposed \acl{JTE}-assisted charge disproportionation reaction as the mechanism behind \ce{Mn} dissolution. In order to unveil the \ce{Mn} dissolution mechanism in \ce{MnO2} cathode cells under \ac{RAZIB} operation conditions, the energetic feasibility for \ce{Mn} vacancy formation was probed in both charged (\ce{MnO2}) and discharged (\ce{ZnMn2O4}) phases of $\alpha$ and $\lambda$ polymorphs of \ce{MnO2} using \acl{DFT}. The formation of a \ce{Mn} vacancy, and consequently the dissolution of \ce{Mn} as \ce{Mn^{2+}_{(aq)}}, was found to be thermodynamically feasible for the $\alpha$-\ce{ZnMn2O4} phase due to the energetically unfavourable \ce{Zn} bent coordination formed during the \ce{Zn^{2+}} intercalation process, indicating that \ce{Mn} dissolution is promoted by an unstable \ce{Zn} coordination environment. The theoretical calculations were then corroborated by \textit{operando} \acl{1HNMR} experiments which captured the \ce{Mn} dissolution occurring throughout the \ac{RAZIB} discharge, with subsequent electrochemical deposition of the \ce{Mn} atoms on the electrode during charge. The combined computational and experimental analysis reveals the critical role of defect energetics and coordination environment in driving active material dissolution, and consequently capacity fade, with the proposed mechanism also relevant for understanding cathode degradation in other intercalating ion battery chemistries. This paper provides fundamental insight into active material dissolution mechanism, offering basis for the development of capacity fade mitigation strategies and discovery of novel cathode materials for prolonged stable battery operation.} \\%The abstrast goes here instead of the text "The abstract should be..."

\end{tabular}

 \end{@twocolumnfalse} \vspace{0.6cm}

  ]
%%%END OF TITLE, AUTHORS AND ABSTRACT%%%

%%%FONT SETUP - please do not change any commands within this section
\renewcommand*\rmdefault{bch}\normalfont\upshape
\rmfamily
\section*{}
\vspace{-1cm}

%%%FOOTNOTES%%%

\footnotetext{\textit{$^{a}$~Department of Materials Science and Engineering, McMaster University, 1280 Main Street West, Hamilton, Ontario L8S 4L8, Canada}}
\footnotetext{\textit{$^{b}$~Department of Chemistry $\&$ Chemical Biology, McMaster University, 1280 Main Street West, Hamilton, Ontario L8S 4L8, Canada}}
\footnotetext{\textit{$^{c}$~Condensed Matter and Statistical Physics Section, The Abdus Salam ICTP, Strada Costiera 11, 34151 Trieste, Italy}}
\footnotetext{\textit{$^{d}$~Salient Energy Inc., 21 McCurdy Avenue, Dartmouth, Nova Scotia B3B 1C4, Canada}}
\footnotetext{\textit{$^{e}$~Department of Chemical Engineering, McMaster University, 1280 Main Street West, Hamilton, Ontario L8S 4L8, Canada}}
\footnotetext{\textit{$^{\ast}$E-mail: miliantc@mcmaster.ca}}

%Please use \dag to cite the ESI in the main text of the article.
%If you article does not have ESI please remove the the \dag symbol from the title and the footnotetext below.
\footnotetext{\dag~Supplementary Information available online.}
%additional addresses can be cited as above using the lower-case letters, c, d, e... If all authors are from the same address, no letter is required

% \footnotetext{\dag~Additional footnotes to the title and authors can be included \textit{e.g.}\ `Present address:' or `These authors contributed equally to this work' as above using the symbols: \ddag, \textsection, and \P. Please place the appropriate symbol next to the author's name and include a \texttt{\textbackslash footnotetext} entry in the the correct place in the list.}

%%%END OF FOOTNOTES%%%

%%%MAIN TEXT%%%%

\acresetall

\section{Introduction}\label{sec:Introduction}

% Introduction to Zinc-ion batteries 
Reliable energy storage solutions are becoming essential to support the ongoing transition from fossil fuels to renewable energy ~\cite{Gallo_RSER_65_2016, Sundararagavan_SE_86_2012, Barton_IEEE_19_2004}. The intermittent nature of solar and wind energy leads to an increase in demand for energy storage infrastructure, where renewable energy can be stored during surplus production and then supplied to the grid during periods of high energy demand~\cite{Gallo_RSER_65_2016, Barton_IEEE_19_2004}. Rechargeable batteries have seen considerable increase in deployment in grid-scale energy storage due to their modular and scalable nature, being the technology behind the majority of new storage capacity gained worldwide~\cite{Schoenfisch_IEA_2023, Gourley_Joule_7_2023, Luo_AppliedEnergy_137_2015}. Currently, \acp{LIB} are the primary chemistry being utilized in rechargeable battery applications at the grid-scale, mainly due to its high energy density (up to 190~W~h~kg$^{-1}$) and cycling life (more than 1000 cycles for 80\% capacity discharge)~\cite{Arteaga_CSRER_4_2017, Diouf_RE_76_2015, Huang_AEM_12_2022}. However, the scarcity and supply chain constraints found for \acs{LIB} component manufacturing (\textit{e.g.}, \ce{Li}, \ce{Co}), alongside the increase in demand for rechargeable batteries, have encouraged researchers to develop alternative battery chemistries to \ac{LIB}~\cite{Hosaka_CR_120_2020, Blanc_J_4_2020, Pan_AEM_11_2021}.

% RAZIB 
One promising beyond lithium battery technology is the \ac{RAZIB}, which is safer, has a lower associated manufacturing cost, and also has a more abundant working ion (\ce{Zn^{2+}}) than \ac{LIB}~\cite{Gourley_Joule_7_2023, Li_AFM_34_2024, Zhou_JCIS_605_2022}. The higher safety and lower cost of manufacturing of \acp{RAZIB} can be attributed to the use of aqueous electrolytes (\textit{e.g.}, \ce{ZnSO4}, \ce{Zn(CF3SO3)2}, \ce{ZnCl2}), thus avoiding the flammable organic electrolytes used in \acp{LIB} which also require purpose-built dry rooms for safe battery fabrication~\cite{Gourley_Joule_7_2023, Blanc_J_4_2020, Huang_CEJ_25_2019}. \acp{RAZIB} are of special interest for implementation at the grid level because, unlike in end-consumer battery applications (\textit{e.g.}, commercial electronics and electric vehicles), the weight and volume limits for grid-scale battery technologies are lower due to their stationary deployment~\cite{Gourley_Joule_7_2023, Li_AFM_34_2024}. The commercial adoption of \acp{RAZIB} would also help to alleviate the current \ac{LIB} supply constraints, as the reliance on \acp{LIB} for grid-scale energy storage applications would decrease~\cite{Gourley_Joule_7_2023, Li_Joule_6_2022, Li_AFM_34_2024}. However the poor cycling performance of currently investigated \ac{RAZIB} cathode materials hampers the commercialization of the technology to become a reality~\cite{Li_AFM_34_2024}.

% MnO2 as cathode material
Transition metal oxides~\cite{Pan_NE_1_2016, Gourley_AMI_17_2025, Tran_SR_11_2021, Kundu_NE_1_2016, Liu_CC_53_2017, He_AS_6_2019, Miliante_JPCC_128_2024, Rubel_JPCC_126_2022}, Prussian blue analogues~\cite{Trocoli_ChemSusChem_8_2015, Zhang_JPS_484_2021}, chalcogenides~\cite{Cheng_AAMI_8_2016, Wang_EF_34_2020, Wang_JPDAP_57_2024}, and organic molecules~\cite{Li_ApplMatInt_14_2022, Gao_ESM_40_2021, Geng_CEJ_446_2022, Espinoza_FB_7_2025, Baker_JES_150_2026, Espinoza_JES_171_2026} are some of the many materials that have already been experimentally investigated as cathode materials for \ac{RAZIB}. While completely different chemical systems, one clear common aspect between all these materials is the significant capacity fade seen at practical cycling rates for grid-scale applications (more than 2 hours per cycle)~\cite{Frazier_NREL_2021, Li_AFM_34_2024, Zampardi_COE_21_2020, Wang_JPDAP_57_2024, Cui_SmartMat_3_2022, Selvakumaran_JMCA_7_2019}. \ce{Mn}-based oxides stand out from all other materials as arguably the most studied cathode materials for \acp{RAZIB}. The popularity of \ce{Mn} oxides can be attributed to the high operating potential (ca.~1.5~V vs Zn/\ce{Zn^{2+}}), eco-friendliness, material abundance, vast number of crystal structures, low price, and wide range of \ce{Mn} valency states~\cite{Selvakumaran_JMCA_7_2019, Zhou_JCIS_605_2022, Xue_JEC_54_2021}. The utilization of \ce{MnO2} as a cathode in \acp{RAZIB} can be dated back to \citeyear{Yamamoto_ICA_117_1986}, when \citet{Yamamoto_ICA_117_1986} reported the cycling of the \ce{Zn}\textbar\ce{ZnSO4}\textbar$\gamma$-\ce{MnO2} aqueous battery system. Since then, different crystal polymorphs of \ce{MnO2} and  modifications to the oxide structure (\textit{e.g.}, oxygen vacancy formation, ion pre-intercalation) have been investigated for improving the cycling performance of \ce{Mn} oxide cathodes~\cite{Zhou_JCIS_605_2022, Azmi_JPS_613_2024, Li_AFM_34_2024}. Multiple studies have previously claimed the development of highly stable \ce{MnO2} cathode materials, achieving more than 1000 cycles with limited capacity fade when cycling at high current densities (higher than or equal to~1~A g${^{-1}}$), consequently having discharges lasting 10 minutes or less~\cite{Tang_ESM_48_2022, Wang_AFM_31_2021, Li_JCIS_676_2024, Xu_Small_19_2023, Wang_CSA_643_2022, Wang_ACSNano_17_2023}. While stability during fast charge/discharge cycles may be appropriate for some applications, having high stability only during fast cycling is not practical for the implementation of \acp{RAZIB} at the grid-scale, since such application requires discharge cycles lasting longer than 2 hours~\cite{Gourley_Joule_7_2023, Frazier_NREL_2021}. 

% MnO2 dissolution and 
The capacity fade seen for \ce{MnO2} cathode materials in \ac{RAZIB} has been attributed to the dissolution of the \ce{Mn} atoms present in the host structure into the aqueous electrolyte as \ce{Mn^{2+}_{(aq)}} during battery discharge \cite{Pan_NE_1_2016, Blanc_J_4_2020, Rubel_JPCC_126_2022, Lee_SR_4_2014, Alfaruqi_EA_276_2018}. The dissolved \ce{Mn} atoms are then electrochemically redeposited into the electrode during the charge cycle~\cite{Wu_EES_13_2020, Rubel_JPCC_126_2022, Zhang_CPBE_260_2023, Liao_ESM_44_2022}. However, the electrodeposition of the \ce{Mn} atoms occurs irregularly on the electrode surface, forming impure phases (\textit{e.g.}, \ce{MnOOH}, layered birnessite (\ce{MnO2}), \ce{ZnMn3O7}), with the initial \ce{MnO2} crystal structure not being completely reestablished in many cases~\cite{Blanc_J_4_2020, Rubel_JPCC_126_2022, Lee_SR_4_2014, Wu_EES_13_2020, Zhang_CPBE_260_2023, Liao_ESM_44_2022}. The electrodeposited \ce{Mn}-containing impure phases, alongside the formation of inert phases (\textit{e.g.}, $\lambda$-\ce{ZnMn2O4}), would then negatively impact the capacity obtained in the following cycles, causing the capacity fade of the \ac{RAZIB}~\cite{Pan_NE_1_2016, Wu_EES_13_2020, Blanc_J_4_2020, Rubel_JPCC_126_2022, Zhang_CPBE_260_2023, Lee_SR_4_2014, Liao_ESM_44_2022}. The higher capacity loss observed during longer-lasting cycling has been attributed to longer time periods spent by the electrode at low potentials, which allows for more \ce{Mn} atoms to dissolve into the electrolyte as \ce{Mn^{2+}_{(aq)}}~\cite{Miliante_JPCC_128_2024}. Therefore, a thorough understanding of the \ce{Mn} dissolution process, and consequently the capacity fade, during cycling rates relevant for grid-scale application is imperative to achieve the commercialization of \acp{RAZIB}. For example, by comprehending the \ce{Mn} dissolution mechanism, it would then be possible to accurately identify materials that are prone to dissolution during operation and develop new strategies for capacity fade mitigation, directly guiding the design of next-generation cathode materials and battery operation strategies. 

% Previous investigations in the dissolution of MnO2 material
% Disproportionation 
The dissolution of \ce{Mn} atoms as \ce{Mn^{2+}_{(aq)}} during \ce{MnO2} cycling has been regularly attributed to a \ac{JTE}-assisted charge disproportionation reaction. First, \ce{Zn^{2+}} ions would intercalate into the \ce{MnO2} structure and form \ce{ZnMn2O4}, with the \ce{Mn} atoms reducing from \ce{Mn}$^{\mathrm{IV}}_{\mathrm{(s)}}$ in \ce{MnO2} to \ce{Mn}$^{\mathrm{III}}_{\mathrm{(s)}}$ in \ce{ZnMn2O4} (Eq.~\eqref{eq:Zn_intercalation}). The \ce{Mn}$^{\mathrm{III}}_{\mathrm{(s)}}$ atoms formed during battery discharge could then go through the \ac{JTE}-assisted charge disproportionation reaction, forming \ce{Mn}$^{\mathrm{IV}}_{\mathrm{(s)}}$, which remain in the solid phase, and soluble \ce{Mn^{2+}_{(aq)}} species, that dissolve in the electrolyte (Eqs.~\eqref{eq:disprop_reaction} and \eqref{eq:disprop_reaction_Zn})~\cite{Lee_SR_4_2014, Tran_SR_11_2021, Siamionau_JPS_523_2022, Cai_CEJ_396_2020, Zhang_NatComms_8_2017, Heo_BS_4_2021, Li_MH_11_2024, Zhang_CPBE_260_2023}.
\begin{eqnarray}
&2\ce{Mn^{IV}O_{2(s)}} +  \ce{Zn^{2+}_{(aq)}} + 2e^{-} \rightarrow \ce{ZnMn}_{2}^{\mathrm{III}}\ce{O4_{(s)}} 
\label{eq:Zn_intercalation}
\\
&2\ce{Mn}_{\mathrm{(s)}}^{\mathrm{III}} \rightarrow \ce{Mn^{2+}_{(aq)}} + \ce{Mn}_{\mathrm{(s)}}^{\mathrm{IV}}
\label{eq:disprop_reaction}
\\
&\ce{ZnMn}_{2}^{\mathrm{III}}\ce{O4_{(s)}}  + 2\ce{H2O} \rightarrow \ce{Zn^{2+}_{(aq)}} + \ce{Mn^{2+}_{(aq)}} + \ce{Mn^{IV}O_{2(s)}} + 4\ce{OH^{-}}
\label{eq:disprop_reaction_Zn}
\end{eqnarray}
The \ac{JTE}-assisted charge disproportionation reaction would be triggered by the formation of \ce{ZnMn2O4} after \ce{Zn^{2+}} intercalation, with unstable \ac{JTE}-active \ce{Mn^{III}O6} octahedra being established in the structure~\cite{Lee_SR_4_2014, Tran_SR_11_2021, Zhang_CPBE_260_2023}. The relationship between the \ac{JTE}-assisted charge disproportionation reaction and capacity fade in batteries was first proposed in the 1980s for the \ce{Li_{\textit{$x$}}Mn_{2}O_{4}} ($1 < x < 2$) cathodes used in \acp{LIB}, with the continuous \ce{Li^{+}} intercalation during discharge being responsible for establishing \ac{JTE}-active \ce{Mn} centres~\cite{Thackeray_ESSL_1_1998, Gummow_SII_69_1994, Thackeray_PSSC_25_1997, Thackeray_MRB_19_1984}. Since then, multiple researchers have questioned the proposed \ac{JTE}-assisted charge disproportionation mechanism for \ce{Mn} dissolution and, consequently, the observed \ac{LIB} capacity fade. For example, \citet{Berg_JMC_9_1999} utilized first-principle calculations to investigate the  stability of fully discharged \ce{Li2Mn2O4} cathodes and found that the \ac{JTE} is crucial for stabilizing the structure, which directly contradicts the proposed mechanism of \ac{JTE}-assisted charge disproportionation. \citet{Wang_JECS_150_2003} demonstrated through the use of rotating ring-disk experiments of a \ce{LiMn2O4} electrode that the concentration of \ce{Mn^{2+}_{(aq)}} in the electrolyte would increase not only after the electrode was discharged but also after being overcharged, a phenomenon that cannot be explained by the proposed \ac{JTE}-assisted charge disproportionation reaction mechanism since the charged phase is not \ac{JTE}-active. Finally, \citet{Bhandari_JECS_164_2016} presented a comprehensive review of the different proposed mechanisms for \ce{Mn} dissolution in spinel cathodes during \ac{LIB} cycling. The authors concluded that the \ac{JTE}-assisted charge disproportionation reaction was not sufficient to explain the capacity fade observed in \ac{LIB} cathodes, since  \ce{Mn} dissolution was demonstrated to occur even for phases that were not \ac{JTE}-active~\cite{Wang_JECS_150_2003, Jang_JECS_143_1996, Takahashi_JPS_136_2004}.

Current \ac{RAZIB} studies still directly rely on the initial proposal by \ac{LIB} studies of a \ac{JTE}-assisted charge disproportionation reaction mechanism to explain the \ce{Mn} dissolution process occurring in \ce{Zn}/\ce{MnO2} aqueous batteries~\cite{Tran_SR_11_2021, Li_MH_11_2024, Siamionau_JPS_523_2022, Heo_BS_4_2021}. However, key differences can be found between the operation of \acp{LIB} and \acp{RAZIB}, such as in electrolyte (non-aqueous vs aqueous), \ce{MnO2} polymorph structure (spinel ($\lambda$) vs tunnelled ($\alpha$, $\beta$, $\gamma$), layered ($\delta$)), and intercalating ion (\ce{Li^{+}} vs \ce{Zn^{2+}}), which puts into question how transferable the findings from one battery chemistry can be to the other, and, to this end, whether or not the \ac{JTE}-assisted charge disproportionation reaction is indeed the key mechanism behind the \ce{Mn} dissolution in \ce{MnO2} cathodes for \acp{RAZIB}.

%%% Vacancy DFT paragraph
The theoretical investigation of the formation of vacancy in solids has been successfully applied in cathodes materials for \acp{LIB} to investigate metallic dissolution and structural transformations during battery operation~\cite{Shiiba_JPCC_119_2015, Ammundsen_JPCB_101_1997, Zhou_ASS_258_2012, Ma_JPD_46_2013, Hu_SSI_347_2020, Leung_CM_29_2017} and after disposal~\cite{AbbaspourTamijani_ASS_515_2020, Huang_JPCC_121_2017, Bennett_EST_52_2018, Hudson_JPCC_127_2023}. For example, \citet{Cai_JCP_158_2023} investigated the dissolution of \ce{Ni}, \ce{Co}, and \ce{Mn} metals at different discharged states of \ce{LiNi_{\textit{x}}Co_{\textit{y}}Mn_{1-\textit{x}-\textit{y}}O2} $(\textit{x}~\geq~0.8)$ cathodes and demonstrated a direct relationship between the transition metal vacancy formation and the structural instability of the cathode material. Also, \citet{He_ATS_3_2020} coupled experimental results and \ac{DFT} calculations to study the impact of thin-film layers deposited onto the \ac{LIB} cathode material on the reduction or acceleration of \ce{Mn} dissolution during cycling, using vacancy formation energy calculations to probe the \ce{Mn} dissolution susceptibility. Despite the significant scientific discoveries on the stability and operation of \ac{LIB} cathode materials achieved through the computational investigation of vacancy formation, limited literature is available on the utilization of similar calculations to study cathode materials under \ac{RAZIB} operation conditions. More importantly, to this date, the formation of \ce{Mn} vacancies in \ce{MnO2} cathodes during \ac{RAZIB} operation has not been pursued, which can prove to be fundamental to uncover the key mechanism behind the \ce{Mn} dissolution process, and support addressing the capacity fade experienced by the battery cells.

% What is done in the article
In this paper, a comprehensive study of the \ce{Mn} dissolution process in \ce{MnO2} cathodes under \ac{RAZIB} operation condition is pursued through a first-principle investigation of the \ce{Mn} vacancy formation. The \ce{Mn} vacancy formation energy was calculated for both neutral and charged vacancy defects in $\alpha$-\ce{MnO2} and $\lambda$-\ce{MnO2} polymorphs, and their respective \ce{ZnMn2O4} discharged phases, capturing the thermodynamic susceptibility for \ce{Mn} dissolution as \ce{Mn^{2+}_{(aq)}} for each evaluated solid phase (\textit{i.e.}, phase stability against \ce{Mn} dissolution). Under \ac{RAZIB} cycling potentials, \ce{Mn} dissolution was found to be thermodynamically favourable to occur in the $\alpha$-\ce{ZnMn2O4} phase, which is formed through  \ce{Zn^{2+}} intercalation into the $\alpha$-\ce{MnO2} cathode during discharge. The \ce{Mn} dissolution mechanism was determined to be connected to a substitution reaction promoted by the unstable bent coordination for \ce{Zn} in $\alpha$-\ce{ZnMn2O4}, with a \ce{Zn} atom establishing an energetically favourable octahedral coordination on the vacant \ce{Mn} site after \ce{Mn} dissolution. The thermodynamic feasibility for \ce{Mn} dissolution during \ac{RAZIB} cycle of an $\alpha$-\ce{MnO2} electrode was also investigated, with \ce{Mn} dissolution demonstrated to be already energetically favourable to occur for a partially discharged $\alpha$-\ce{Zn_{x}Mn2O4} ($x$~$<$~0.5) phase due the unstable \ce{Zn} coordination environment. Finally, \textit{operando} \acl{1HNMR} experiments were utilized to quantify the concentration of \ce{Mn^{2+}_{(aq)}} in the electrolyte during \ac{RAZIB} cycling, with the continuous \ce{Mn} dissolution from the $\alpha$-\ce{MnO2} electrode revealed to occur throughout the discharge process. After uncovering the unstable \ce{Zn} coordination as the source for \ce{Mn} dissolution in $\alpha$-\ce{MnO2} electrodes, potential strategies for capacity fade mitigation centred around altering the intercalating ion coordination environment were proposed, such as the inclusion of foreign atoms inside the tunnelled structure and incorporation of alloying/doping atoms in the \ce{Mn} sites. The proposed dissolution mechanism can be regarded as an universally applicable degradation mechanism for battery cathodes that operate via ion intercalation, since the formation of an unstable ion coordination environment that will promote the dissolution is independent of battery chemistry. Therefore, the methodology and results presented in this study provide crucial insight into cathode material degradation, which will ultimately support addressing the battery capacity fade occurring during cycling and expedite the proposal of novel cathode materials that are stable under battery operation conditions.

\section{Methods}\label{sec:Methods}
% Computational methods

\subsection{Computational Methods}\label{sec:CompMethods}

Plane-wave \ac{DFT} calculations with the \Ac{PBE} exchange-correlational functional~\cite{Perdew_PRL_77_1996} were carried out utilizing the \ac{VASP}~\cite{Kresse_PRB_47_1993, Kresse_CMS_6_1996, Kresse_PRB_54_1996} (version 6.4.3). Projector augmented wave pseudopotentials~\cite{Kresse_PRB_59_1999} for \ce{Mn}, \ce{O} and \ce{Zn} were used, with 3p$^{6}$4s$^{2}$3d$^{5}$, 2s$^{2}$2p$^{4}$, and 4s$^{2}$3d$^{10}$ valence electrons respectively considered. The plane-wave cutoff energy was set equal to the highest cutoff energy from all considered pseudopotentials, resulting in a 400~eV cutoff energy being employed. The long-range van der Waals interactions were accounted for with the use of the DFT-D3 method proposed by Grimme, with the Becke-Johnson damping function~\cite{Grimme_JCP_132_2010, Grimme_JCC_32_2011, Becke_JCP_123_2005, Johnson_JCP_123_2005, Johnson_JCP_124_2006}. The Hubbard \textit{U} correction for the highly correlated \textit{d} electrons of \ce{Mn} was considered in all calculations following the implementation of \citet{Dudarev_PRB_57_1998}, with the \textit{U}$_{\text{eff}}$ value equal to 3.9~eV as proposed by \citet{Jain_PRB_84_2011}. Spin polarization with ferromagnetic ordering was also considered for all oxides studied. Antiferromagnetic ordering was also investigated, however overall higher energy for the structures were obtained. The higher energy found for antiferromagnetic structures can be explained by the use of a high \textit{U}$_{\text{eff}}$ in our calculations, as explained by \citet{Crespo_PRB_88_2013}. The vacancy formation energy in $\alpha$-\ce{MnO2} was calculated to be only of approximately 0.5~eV higher when considering antiferromagnetic ordering, a difference in magnitude which ultimately should not change the conclusions of this study.

A $\Gamma$-centred \textit{k}-mesh with a density of 20~subdivisions per $\text{\AA}^{-1}$ was considered for sampling the Brillouin zone. The choice of \textit{k}-mesh was verified from \ac{SC} calculations of the relaxed structures with a denser \textit{k}-mesh, which demonstrated negligible variations in the calculated defect formation energy. Full atomic position and cell parameter relaxation was performed for all defect-free systems, with convergency tolerances of 10$^{-7}$~eV and 10$^{-2}$~eV~$\text{\AA}^{-1}$ respectively considered for the energy change in consecutive \ac{SC} loops and for the norm of ionic forces. Only changes in the atomic positioning were allowed for the relaxation of cells containing point defects, with the same convergency tolerances being considered. A Gaussian smearing of 0.01~eV was employed in all calculations.

% Structures
The $\alpha$-\ce{MnO2} and $\lambda$-\ce{MnO2} polymorphs were selected as representative materials for the investigation of the \ac{VMn} formation in \ce{MnO2} cathode materials. $\alpha$-\ce{MnO2} is a prominent and actively researched cathode material for \acp{RAZIB}. The $\alpha$-\ce{MnO2} structure is composed of 2$\times$2 tunnels that facilitate \ce{Zn^{2+}} intercalation during battery discharge, with the structure consequently having considerably high practical capacity (greater than 200~mAh~g$^{-1}$)~\cite{Selvakumaran_JMCA_7_2019, Zhou_JCIS_605_2022}. However, severe capacity fade has been reported in $\alpha$-\ce{MnO2} cathode materials due to the dissolution of \ce{Mn} atoms from the structure during battery cycling~\cite{Lee_SR_4_2014, Zhang_CPBE_260_2023, Wu_Small_14_2018, Selvakumaran_JMCA_7_2019, Zhou_JCIS_605_2022, Miliante_JPCC_128_2024}. On the other hand, considerably less research attention has been given to $\lambda$-\ce{MnO2} as a cathode for \acp{RAZIB} due to the limited ion migration space in its spinel structure, which negatively impacts the possibility of \ce{Zn^{2+}} intercalation~\cite{Zhang_JEC_82_2023, Zhou_JCIS_605_2022}. However, $\lambda$-\ce{MnO2} is the only \ce{MnO2} polymorph for which the synthesis of the \ce{Zn}-containing discharged phase \ce{ZnMn2O4} has been experimentally confirmed~\cite{Patra_PB_572_2019, Lobo_JMS_27_2016, Courtel_EA_71_2012}. Therefore, a more complete analysis of the impact of \ce{Zn} on the formation of \ac{VMn} in \ce{MnO2} can be expected by including both the $\alpha$ and $\lambda$ polymorphs in our study.

The unit cell structures for $\alpha$-\ce{MnO2} (mp-19395, 2$\times$2$\times$3 supercell), $\lambda$-\ce{MnO2} (mp-25275, 2$\times$2$\times$2 supercell), and $\lambda$-\ce{ZnMn2O4} (mp-18751, 2$\times$2$\times$1 supercell) were obtained from the Materials Project database~\cite{Jain_AM_1_2013}, with the corresponding supercells created from the respective unit cells. The relaxed supercell lattice parameters were verified to be in agreement with the experimental unit cell lattice parameters~\cite{Kijima_JSSC_177_2004, Hunter_JSSC_39_1981, Menaka_BMS_32_2009}, as shown in Table~S1. It was necessary to create a supercell from each structure in order to minimize the impact of the system periodicity during the investigation of the \ac{VMn} formation in the bulk materials. Unfortunately, the crystallographic structure of $\alpha$-\ce{ZnMn2O4} has not been experimentally resolved; therefore, the $\alpha$-\ce{ZnMn2O4} structure had to be obtained directly from first principles by incorporating \ce{Zn} atoms into the $\alpha$-\ce{MnO2} tunnelled structure. The optimal \ce{Zn} atom positioning inside $\alpha$-\ce{MnO2} was determined by comparing the energy from fully relaxed unit cell-sized $\alpha$-\ce{ZnMn2O4} structures with different \ce{Zn} placements investigated. The \ce{Zn} placement in the lowest energy $\alpha$-\ce{ZnMn2O4} unit cell structure was then replicated in the $\alpha$-\ce{MnO2} supercell, thus obtaining the initial $\alpha$-\ce{ZnMn2O4} supercell structure before relaxation.  

% Defect formation energy calculation 
A general chemical equation for the formation of a single \acf{VMn} in \ce{MnO2} at a charge state \textit{q} with the dissolution of the \ce{Mn} atom as \ce{Mn^{2+}_{(aq)}} can be written as
\begin{equation}
\ce{Mn^{IV}O_{2(s)}} \rightarrow \ce{Mn_{1-\epsilon}\vrectangleA^{\textit{q}}_{\epsilon}O_{2(s)}} + \epsilon\ce{Mn^{2+}_{(aq)}} + \epsilon(2+q)\mathrm{e^{-}},
\label{eq:MnO2_vacancy}
\end{equation}
where \ce{MnO2_{(s)}} and \ce{Mn_{1-\epsilon}\vrectangleA^{\textit{q}}_{\epsilon}O2_{(s)}} are the structures for \ce{MnO2} without and with the \ac{VMn}, such that $\epsilon\ce{Mn^{2+}_{(aq)}}$ corresponds to only 1 \ce{Mn} atom. The \ce{\vrectangleA^{\textit{q}}_{\epsilon}} notation in \ce{Mn_{1-\epsilon}\vrectangleA^{\textit{q}}_{\epsilon}O2_{(s)}} indicates the presence of $\epsilon$  vacancies with charge $\textit{q}$ per formula unit of the host structure. From Eq.~\eqref{eq:MnO2_vacancy}, it can be seen that the formation of a \ac{VMn} will be an electrochemically oxidative ($q>-2$), an electrochemically reductive ($q<-2$), or a non-electrochemical ($q=-2$) process depending on the charge \textit{q} considered for the formed defect. In this article, only the formation of a neutral defect ($q=0$) and a negatively charged defect ($q=-2$) will be investigated. Investigating both defect types allows for the analysis of a scenario where the surplus charge from the \ac{VMn} formation is directed to the battery system ($q=0$) or accumulated in the cathode material ($q=-2$). An analogous chemical equation can also be established for \ce{ZnMn2O4}:
\begin{equation}
\ce{ZnMn}_{2}^{\mathrm{III}}\ce{O4_{(s)}} \rightarrow \ce{ZnMn_{2-\epsilon}\vrectangleA^{\textit{q}}_{\epsilon}O_{4(s)}} + \epsilon\ce{Mn^{2+}_{(aq)}} + \epsilon(2+q)\mathrm{e^{-}}.
\label{eq:ZnMn2O4_vacancy}
\end{equation} 

Equations~\eqref{eq:MnO2_vacancy} and \eqref{eq:ZnMn2O4_vacancy} outline the chemical reactions for the \ce{Mn} dissolution as \ce{Mn^{2+}_{(aq)}} through the formation of a \ce{Mn} vacancy in the host structures. The energetic susceptibility for \ce{Mn} dissolution from each material can then be determined through the calculation of the energy variation associated with the vacancy formation reactions (Eqs.~\eqref{eq:MnO2_vacancy} or \eqref{eq:ZnMn2O4_vacancy}), with this parameter being known as the defect formation energy ($E_{\mathrm{d}}$). Following the reaction paths for Eqs.~\eqref{eq:MnO2_vacancy} and \eqref{eq:ZnMn2O4_vacancy}, a negative $E_{\mathrm{d}}$ ($E_{\mathrm{d}}$~$<$~0) indicates that the \ce{Mn} vacancy formation, and consequently the \ce{Mn} dissolution, will be energetically favourable to occur. Conversely, a positive $E_{\mathrm{d}}$ ($E_{\mathrm{d}}$~$>$~0) will point to the \ce{Mn} vacancy formation being unfeasible to occur, with the \ce{Mn} atom being predicted to remain in the electrode. Apart from determining if the \ce{Mn} dissolution is energetically possible, $E_{\mathrm{d}}$ will also capture the degree of instability for the phase, with a more negative $E_{\mathrm{d}}$ indicating a greater drive for the \ce{Mn} dissolution reaction (\textit{i.e.}, higher material instability).

The $E_{\mathrm{d}}$ for a single \ac{VMn} in a charge state \textit{q} ($\mathrm{V}^{~q}_{\ce{Mn}}$) can then be calculated by~\cite{Zhang_PRL_67_1991, Freysoldt_RMP_86_2014, Rubel_JPCC_128_2024, Rong_JPCL_6_2015} 
\begin{equation}
\begin{gathered}
E_{\mathrm{d}}(\mathrm{V}^{~q}_{\ce{Mn}},{\mathrm{\Delta}}E_{\mathrm{F}},\acs{phiSHEMn}) = E_{\mathrm{tot}}(\zeta(\mathrm{V}^{~q}_{\ce{Mn}}))- E_{\mathrm{tot}}(\zeta({\mathrm{host})}) + \mu (\ce{Mn^{2+}_{(aq)}}) \\ + q\{E_{\mathrm{VBE}}(\zeta({\mathrm{host}})) + {\mathrm{\Delta}}E_{\mathrm{F}}\}  
+ E_{\mathrm{corr}} - (2+q)\acs{phiSHEMn},
\end{gathered}
\label{eq:FormEnergyDefect}
\end{equation}
where $E_{\mathrm{tot}}(\zeta)$ is the \ac{DFT} total energy for the system $\zeta$, $\zeta(\mathrm{V}^{~q}_{\ce{Mn}})$ refers to the system containing the vacancy, $\zeta(\mathrm{host})$ is the system where the defect is being formed (\textit{i.e.}, pristine material), $\mu(\ce{Mn^{2+}_{(aq)}})$ is the chemical potential of \ce{Mn^{2+}_{(aq)}} in the condition of interest, $E_{\mathrm{VBE}}(\zeta({\mathrm{host}}))$ is the \ac{VBE} energy obtained from the calculation of the host system, ${\mathrm{\Delta}}E_{\mathrm{F}}$ accounts for the position of the V$^{~q}_{\ce{Mn}}$ system Fermi energy with respect to the $E_{\mathrm{VBE}}(\zeta({\mathrm{host}}))$, $E_{\mathrm{corr}}$ is the energy correction term for charged defect calculations, and $\phi_{\mathrm{SHE}}$ is the \acl{phiSHEMn}. The applied corrections were: (i) a potential alignment between the charged system and the neutral host~\cite{Lany_MSMS_17_2009}, and (ii) the \ac{FNV} correction for the spurious interactions between charged defects in finite-sized supercells~\cite{Freysoldt_PRL_102_2009}. The correction scheme proposed by \citet{Freysoldt_PRL_102_2009} allowed our energy correction to account for the anisotropy of the dielectric constant for the charged systems, which were calculated through density functional perturbation theory. The \ac{FNV} correction was applied for the charged vacancy results in $\alpha$-\ce{MnO2} and $\lambda$-\ce{MnO2}, since only for these two structures was the charged defects determined to be localized. 

% Chemical potentials
The chemical potential $\mu(\ce{Mn^{2+}_{(aq)}})$ can be calculated from 
\begin{equation}
\mu(\ce{Mn^{2+}_{(aq)}}) = \mu^{\circ}(\ce{Mn}) + {\mathrm{\Delta}}\mu(\ce{Mn^{2+}_{(aq)}}),
\label{eq:ChemicalPotentialTheta}
\end{equation}
where $\mu^{\circ}(\ce{Mn})$ is the standard chemical potential of \ce{Mn}, and ${\mathrm{\Delta}}\mu(\ce{Mn^{2+}_{(aq)}})$ is the variation in the chemical potential of \ce{Mn} between the standard thermodynamic condition and the condition of interest. $\mu^{\circ}(\ce{Mn})$ can be approximated to the \ce{DFT} calculated total energy per atom of \ce{Mn} in its most stable polymorph ($E_{\mathrm{tot}}(\ce{Mn^{0}_{(s)}})$), since \ce{Mn} is solid at standard conditions~\cite{Zhang_PRL_67_1991}. The tetragonal ferrimagnetic structure of $\alpha$-\ce{Mn} was considered here for the calculation of $E_{\mathrm{tot}}(\ce{Mn^{0}_{(s)}})$ following the theoretical results obtained by \citet{Hobbs_PRB_68_2003} on the investigation of magnetism in $\alpha$-\ce{Mn}. Different magnetic orderings were investigated, with the ferrimagnetic system (0.45~$\textit{\micro}_{B}$~f.u.$^{-1}$) achieving the lowest total energy. The condition of interest for \ce{Mn} is \ce{Mn^{2+}_{(aq)}} ions at the concentration found for \ac{RAZIB} electrolytes, given that the dissolution of \ce{Mn} atoms from the cathode material into the electrolyte is being investigated. Thus, ${\mathrm{\Delta}}\mu(\ce{Mn^{2+}_{(aq)}})$ can be calculated by
\begin{equation}
{\mathrm{\Delta}}\mu(\ce{Mn^{2+}_{(aq)}}) = \mathrm{\Delta}G^{\circ}(\ce{Mn^{2+}_{(aq)}}) + \mathrm{R}T\mathrm{ln}([\ce{Mn^{2+}_{(aq)}}]),
\label{eq:DeltaMuMn2+}
\end{equation}
where $\mathrm{\Delta}G^{\circ}(\ce{Mn^{2+}_{(aq)}})$ is the standard Gibbs energy of \ce{Mn^{2+}_{(aq)}} (-2.36~eV~\cite{Pourbaix1974_Mn}), R is the ideal gas constant (8.617~10$^{-5}$~eV~K$^{-1}$~atom$^{-1}$), \textit{T} is the temperature at the condition of interest (298~K), and [\ce{Mn^{2+}_{(aq)}}] is the concentration of \ce{Mn^{2+}_{(aq)}} in the \ac{RAZIB} electrolyte. Previous studies have reported improvements in the cycling capacity and capacity retention for battery cells with \ce{MnO2} cathodes through the addition of \ce{Mn^{2+}_{(aq)}} into the electrolyte at concentrations around 0.1~M \cite{Chamoun_ESM_15_2018, Cui_JPS_579_2023}. Following the adoption by researchers of \ce{Mn^{2+}_{(aq)}} electrolyte additives on the development of \ce{Mn}-based \ac{RAZIB} cathodes, a concentration of \ce{Mn^{2+}_{(aq)}} in the electrolyte equal to 0.1~M was considered in our calculations.

The \ac{RAZIB} cathode operation potential is commonly reported with respect to the \ce{Zn} stripping/plating reaction occurring on the battery anode (\ce{Zn}/\ce{Zn^{2+}} redox couple). Therefore, the \ac{phiZnMn} can be calculated from \acs{phiSHEMn} by
\begin{equation}
\acs{phiZnMn} = \acs{phiSHEMn} - \left\{\phi^{\circ}_{\mathrm{SHE}}(\ce{Zn}/\ce{Zn^{2+}}) +  \frac{RT}{z}\mathrm{ln}([\ce{Zn^{2+}_{(aq)}}])\right\}, 
\label{eq:phiZnMn}
\end{equation}
where $\phi^{\circ}_{\mathrm{SHE}}(\ce{Zn}/\ce{Zn^{2+}})$ is the \acl{phiSHEZnstd} (-0.762~V~vs~SHE~\cite{Rumble_CRC_ElectSeries_2023}), $[\ce{Zn^{2+}_{(aq)}}]$ is the concentration of \ce{Zn^{2+}_{(aq)}} in the electrolyte (considered as 1~M~\cite{Miliante_JPCC_128_2024, Liu_AEL_6_2021}), and \textit{z} is the number of electrons associated with the \ce{Zn}/\ce{Zn^{2+}} redox couple reaction (\textit{z}~=~2). Finally, the equation for \ac{EdVMn} can be written as
\begin{equation}
\begin{gathered}
E_{\mathrm{d}}(\mathrm{V}^{~q}_{\ce{Mn}},{\mathrm{\Delta}}E_{\mathrm{F}},\acs{phiZnMn}) = E_{\mathrm{tot}}(\zeta(\mathrm{V}^{~q}_{\ce{Mn}}))- E_{\mathrm{tot}}(\zeta({\mathrm{host}})) \\ + \left\{\mu^{\circ}(\ce{Mn}) + \mathrm{\Delta}G^{\circ}(\ce{Mn^{2+}_{(aq)}}) + \mathrm{RTln}([\ce{Mn^{2+}_{(aq)}}])\right\} \\ + q\{E_{\mathrm{VBE}}(\zeta({\mathrm{host}})) + {\mathrm{\Delta}}E_{\mathrm{F}}\}  + E_{\mathrm{corr}} \\ - (2+q)\{\acs{phiZnMn} +  \phi^{\circ}_{\mathrm{SHE}}(\ce{Zn}/\ce{Zn^{2+}})\},
\end{gathered}
\label{eq:FormEnergyDefect_final}
\end{equation}

A fork~\cite{Rubel_github_PyDEFFork} of the PyDEF 2~\cite{Stoliaroff_JCC_39_2018, Pean_CPL_671_2017} code, which added compatibility with \ac{VASP}~6, was used for the calculation of the formation energy of the defects. The CoFFEE code~\cite{Naik_CPC_226_2018} was used for the calculation of the \ac{FNV} correction, with the correction value incorporated into the results from PyDEF.

\subsection{Experimental Methods}\label{sec:ExpMethods}

The pristine $\alpha$-\ce{MnO2} active powder was supplied by Vibrantz Technologies and used without modification. \Ac{SEM} analysis of the pristine powder was performed and is presented in Fig.~S1, with an average primary particle size of ca.~5~$\micro$m being seen. The cathodes were prepared in house by first creating a slurry 50$\mathrm{\%}$ $\alpha$-\ce{MnO2}, 40$\mathrm{\%}$ conductive carbon black and 10$\mathrm{\%}$ \ac{PTFE} binder dissolved in a 2:1 mixture of Millipore Type I Ultra-pure water and isopropanol. This slurry was mixed using a planetary mixer (Mazerustar KK-250S) for 10~min at 900~rpm until a smooth and flowable slurry was achieved. It was then casted along the surface of the carbon paper current collector (AvCarb P50, AvCarb Materials Solutions) using a doctor blade with a 250~$\micro$m gap, then placed in an oven at 100~$\degree$C for a minimum of 3~h to allow for the water and isopropanol to fully evaporate. The resulting average active material loading was 0.80~mg~cm$^{-2}$.

The experimental \ac{RAZIB} cell was assembled using rectangular $\alpha$-\ce{MnO2} electrodes cut using a custom die of dimensions 1.9×3.0~cm with 2~mm radius corners. A similar die of dimensions 2.0×3.1~cm was used to cut the glass microfiber separator (Whatman GF/D) and imprint the desired shape on to \ce{Zn} metal foil, which was then cut out using scissors. The electrode assembly was placed into the bottom side of a cartridge cell designed for use in \textit{in-situ} \acs{NMR} experiments, as described previously~\cite{Sanders_Carbon_2022_189, Sanders_JACS_2023_145}, with 710~$\micro$L of aqueous 1~M~\ce{ZnSO4} electrolyte being added. The cell was closed and rested overnight so that the cell is uniformly wetted prior to cycling. All \ac{RAZIB} tests were conducted utilizing the cartridge cell setup just described.

The \ce{Mn} dissolution process occurring during \ac{RAZIB} cycling was probed through an analysis of the \ac{Mn2+conc} in the electrolyte captured by \textit{operando} \acf{1HNMR} experiments. All \ac{1HNMR} spectroscopy measurements were carried out at an external magnetic field strength of 7.05~T, nominally 300.33~MHz for $^{1}$H. The electrochemical cell was placed in a parallel plate resonator tuned to 300~MHz~\cite{Aguilera_MRL_2023_3}, which was connected to a Bruker MicWB40 probe. A Bruker Avance IIIHD spectrometer was used for collection of \ac{1HNMR} data and all spectra were processed and plotted using Topspin. \textit{Operando} \ac{1HNMR} spectra were collected during cell cycling using a pulse-acquire sequence with a pulse tip angle of $\beta~\approx16\degree$ and a repetition time of 0.9375~s. Sixty-four scans were acquired per 1D acquisition, resulting in a data collection time of 60~s for each \textit{operando} \ac{1HNMR} spectrum. \textit{In situ} measurements of the \ce{H2O} $T_{1}$ relaxation time were acquired using an inversion recovery sequence and analyzed in Topspin using a two-component fit, as described recently~\cite{Sanders_CM_38_2026}. The resulting fit was then used to determine the \ac{Mn2+conc} in the cell electrolyte~\cite{Sanders_CM_38_2026}. 

The \textit{in situ} cell was placed in the \ac{1HNMR} probe and attached to a Gamry Interface 1010E potentiostat for electrochemical cycling using the same setup as described previously~\cite{Sanders_JACS_2023_145, Sanders_CM_38_2026, Sadighi_ChemRxiv_2025}. The cell was discharged to 1.1~V using a current of 0.13~mA~cm$^{-2}$ (0.16~mA~mg$^{-2}$), at which point the cell was rested for one hour under open circuit conditions so that the electrolyte could return to equilibrium. The \textit{operando} \ac{1HNMR} experiments were then halted and the \textit{in situ} determination of \ac{Mn2+conc} by analysis of the solvent T$_{1}$ was carried out. After completion of the T$_{1}$ determination, \textit{operando} \ac{1HNMR} measurements were simultaneously started while beginning to charge the cell. The cell was charged to 1.8~V using a current of 0.13~mA~cm$^{-2}$ and again allowed to rest for one hour prior to \ac{Mn2+conc} measurement. The cell was then cycled symmetrically three times using the same currents without \ac{OCV} holds and subsequently discharged, rested at open circuit for \ac{1HNMR} experiments, then charged similarly as before for \ac{1HNMR} measurements after the fifth discharge and charge steps. This was repeated to obtain \textit{in situ} T$_{1}$ measurements for cycles 10, 15, and 20 interleaved with \textit{operando} \ac{1HNMR} measurements during cycling. The percentage of \ce{Mn} dissolved from the electrode was determined with respect to the initial mass of \ce{MnO2} in the electrode, and then calculated by dividing the obtained \ac{Mn2+conc} from the \ac{1HNMR} measurement by the maximum \ac{Mn2+conc} possible if all the active material had dissolved into the electrolyte.

\section{Results and Discussion}\label{sec:ResultsDiscussion}

\begin{figure}
\centering
\includegraphics[width=0.35\textwidth]{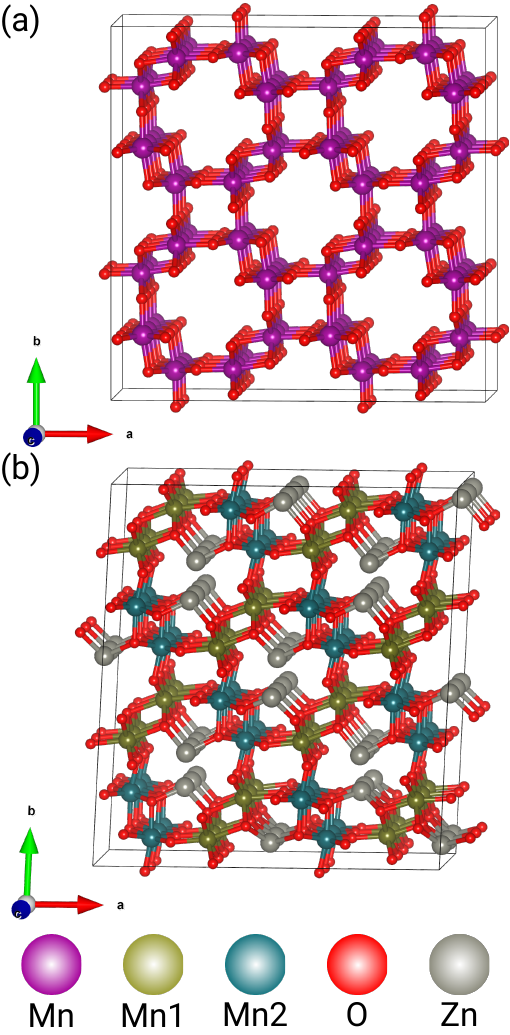}
\caption{Relaxed crystal structures of bulk (a) $\alpha$-\ce{MnO2} and (b) $\alpha$-\ce{ZnMn2O4}. As described in the text, the $\alpha$-\ce{ZnMn2O4} \ce{Mn} atoms have been classified as \ce{Mn}1 (olive) or \ce{Mn}2 (teal) depending on the orientation of their \ce{MnO6} octahedral with respect to the bent coordination of the \ce{Zn} atoms.}
\label{fig:InitialStructures}
\end{figure}

% Structure defect-free \alpha systems
The supercell structures of $\alpha$-\ce{MnO2} and $\alpha$-\ce{ZnMn2O4} after relaxation are shown in Fig.~\ref{fig:InitialStructures}. The $\alpha$-\ce{MnO2} structure is formed by 2$\times$2 tunnels, with \ce{MnO6} octahedra in two different orientations forming the tunnels (Fig.~\ref{fig:InitialStructures}a). In the case of $\alpha$-\ce{ZnMn2O4}, the intercalated \ce{Zn^{2+}} ions established themselves within the tunnels, forming a bent coordination with the \ce{O} atoms (\ce{ZnO2}), with the characteristic 2$\times$2 tunnels from $\alpha$-\ce{MnO2} being preserved (Fig.~\ref{fig:InitialStructures}b). The \ce{Mn} atoms in $\alpha$-\ce{MnO2} can all be considered crystallographically equivalent due to their symmetry within the structure, despite the difference in the octahedral orientations in the tunnels. However, for the discharged structure $\alpha$-\ce{ZnMn2O4}, the \ce{Mn} atoms are not crystallographically equivalent anymore. The symmetry loss for the \ce{Mn} atoms in $\alpha$-\ce{ZnMn2O4} is a result of the \ce{ZnO2} bent coordination formed after the \ce{Zn^{2+}} intercalation, which caused two crystallographically different \ce{Mn} sites to be established. Therefore, it becomes necessary to differentiate the \ce{Mn} atoms present in $\alpha$-\ce{ZnMn2O4} with respect to their orientation to the intercalated \ce{Zn} atoms. The \ce{Mn} atoms that have their \ce{MnO6} octahedral orientation in the tunnel parallel to the \ce{ZnO2} bent coordination are classified as \ce{Mn}1 (olive-coloured atoms in Fig.~\ref{fig:InitialStructures}b). Meanwhile, the \ce{Mn} atoms with \ce{MnO6} octahedral orientation perpendicular to the \ce{ZnO2} coordination are labelled \ce{Mn}2 (teal-coloured atoms in Fig.~\ref{fig:InitialStructures}b). A complete analysis regarding the \ac{VMn} formation in $\alpha$-\ce{ZnMn2O4} can then be expected after distinguishing between the two different \ce{Mn} sites, as different results could be obtained for each site due to their crystallographic difference. The relaxed supercell structures of $\lambda$-\ce{MnO2} and $\lambda$-\ce{ZnMn2O4} can be seen in Fig.~S2 in the \ac{SI}, with both materials retaining their cell symmetry and crystallographic equivalency for the \ce{Mn} atom sites after relaxation.

\begin{figure*}
\centering
\includegraphics[width=0.85\textwidth]{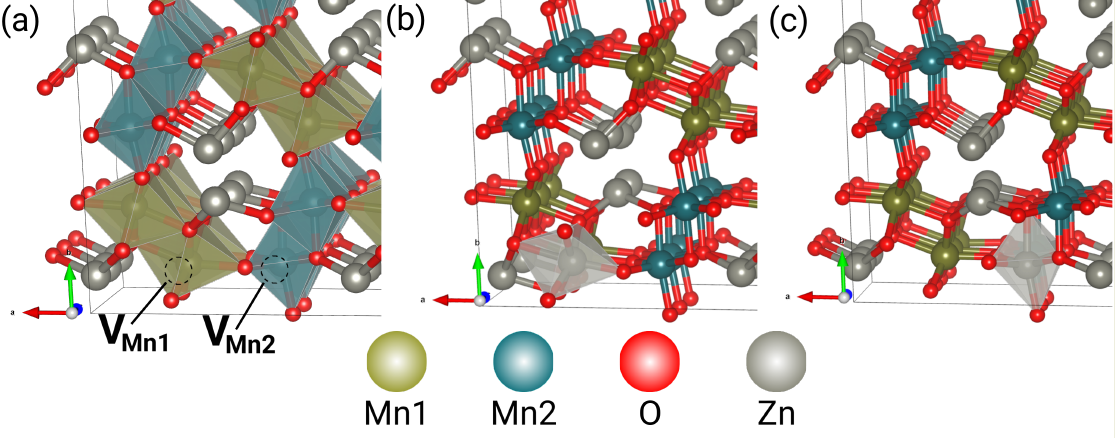}
\caption{(a) \ac{VMn1} and \ac{VMn2} site position in the $\alpha$-\ce{ZnMn2O4} crystal structure indicated by dashed-contoured circles. Relaxed structures of (b) $\alpha$-\ce{ZnMn2O4(\ac{VMn1})} and (c) $\alpha$-\ce{ZnMn2O4(\ac{VMn2})} for neutral charge defects ($q~=~0$), with the \ce{ZnO6} octahedra established after relaxation highlighted as a polyhedron. Only a zoomed-in region of the respective simulated $\alpha$-\ce{ZnMn2O4}(\acs{VMn}) cells is presented here to facilitate the visual analysis of the \ac{VMn} positioning and atomic displacements during the relaxation of the \ce{Mn}-vacant structures.}
\label{fig:VacancyStructures}
\end{figure*}

The positions of the \ac{VMn1} and \ac{VMn2} sites in $\alpha$-\ce{ZnMn2O4} are presented in Fig.~\ref{fig:VacancyStructures}a, while in Figs.~\ref{fig:VacancyStructures}b~and~\ref{fig:VacancyStructures}c the relaxed $\alpha$-\ce{ZnMn2O4} structures for the neutral ($q$=0) \ac{VMn1} and \ac{VMn2} defects can be respectively seen. It was found from the atomic relaxation of $\alpha$-\ce{ZnMn2O4}(\ac{VMn1}) that the lowest energy structure had the \ac{VMn1} site occupied by a previously neighbouring \ce{Zn} atom (Fig.~\ref{fig:VacancyStructures}b). The \ce{Zn} atom now on the \ac{VMn1} site establishes an octahedral coordination with the oxygen atoms (\ce{ZnO6}), similarly to the \ce{Mn}1 atom that formed the vacancy. The formation of \ce{ZnO6} octahedral coordination at the \ac{VMn} position was also observed in the relaxed $\alpha$-\ce{ZnMn2O4}(\ac{VMn2}) structure (Fig.~\ref{fig:VacancyStructures}c). The structural transformation undergone by both $\alpha$-\ce{ZnMn2O4}(\ac{VMn}) structures can be understood as a \ce{Zn} substitution reaction, with the previously intercalated \ce{Zn} atom replacing the \ce{Mn} atom in the octahedral after the \ce{Mn} vacancy is formed and \ce{Mn} is dissolved into the electrolyte. The shift in \ce{Zn} coordination during the relaxation of the vacant structure potentially indicates that the bent \ce{Zn} coordination established after intercalation is energetically unstable, with a more stable octahedral coordination only structurally possible to be formed after the \ce{Mn} atom dissolution. A more detailed analysis of the impact of the structural transformations undergone by $\alpha$-\ce{ZnMn2O4}(\ac{VMn}) on the propensity for \ce{Mn} dissolution from the structure is presented later, alongside the discussion of the calculated \ac{EdVMn} for the material. In the case of $\alpha$-\ce{MnO2}(\ac{VMn}), $\lambda$-\ce{MnO2}(\ac{VMn}), and $\lambda$-\ce{ZnMn2O4}(\ac{VMn}), negligible structural transformations were observed after relaxation (see Figs.~S3b, S4b and S4d in the \ac{SI}).

%%%%% Discussion about the formation of neutral defects

First, the \ac{EdVMn0} ($q=0$)  was calculated for all the materials considering an electrode potential equal to 0~V~vs~SHE (\acs{phiZnMn}~=~0.762~V~vs~\ce{Zn}/\ce{Zn^{2+}} in Eq.~\eqref{eq:FormEnergyDefect_final}), and is shown in Table~\ref{tab:VacResults}. An electrode potential of 0~V~vs~SHE was chosen here because it allows for a direct comparison to previously reported \ac{EdVMn0} results for \ce{Mn} oxides. The high \ac{EdVMn0} values seen for both \ce{MnO2} polymorphs (7~eV or higher) display how energetically unfavourable the \ac{VMn} formation and subsequent dissolution as \ce{Mn^{2+}_{(aq)}} would be in both structures. Similar \ce{Mn} vacancy formation calculations were previously performed for different \ce{MnO2} polymorphs, with formation energies greater than 10~eV being reported \cite{Young_CM_27_2015, Song_CMS_138_2017}. The disparity between our results and previous reports can be ascribed to the difference in chemical environment considered for the \ce{Mn} atom in the \ac{EdVMn0} calculation, which was here taken as \ce{Mn^{2+}_{(aq)}} (see Eqs.~\eqref{eq:FormEnergyDefect}, \eqref{eq:ChemicalPotentialTheta} and \eqref{eq:DeltaMuMn2+}). For example, \citet{Young_CM_27_2015} calculated a \ac{EdVMn0} in $\alpha$-\ce{MnO2} to be of approximately 11~eV when considering a \ce{Mn} chemical environment that of \ce{Mn^{0}_{(s)}} (\ce{Mn}-rich environment) and without performing defective cell relaxation. The \ac{EdVMn0} result by \citet{Young_CM_27_2015} can then be adjusted to be with respect to \ce{Mn^{2+}_{(aq)}} (Eqs.~\eqref{eq:ChemicalPotentialTheta} and \eqref{eq:DeltaMuMn2+}), resulting in a \ac{EdVMn0} of approximately 8.6~eV, which is similar to our results for \ce{MnO2} polymorphs without relaxation (Table~\ref{tab:VacResults}). Overall, all studies agree that the formation of a \ac{VMn} is energetically highly unfavourable, independent of the \ce{MnO2} polymorph considered; thus, it is expected that the \ce{Mn} atom will remain in the \ce{MnO2} lattice. 

\newcolumntype{M}[1]{>{\centering\arraybackslash}m{#1}}
% \newcolumntype{N}{@{}m{0pt}@{}}
\setlength\extrarowheight{6pt}

\begin{table*}
\caption{Results for the \acf{EdVMn0} and \acf{EdVMn-2CBE} for the \ce{MnO2} and \ce{ZnMn2O4} polymorphs, alongside the supporting parameters: \acf{dErelax} and \acf{EdVMn0relax} (described in the text). All values are in eV. At the conditions investigated in this paper, an electrode potential of~0.762~V~vs~\ce{Zn}/\ce{Zn^{2+}} is equivalent to~0~V~vs~SHE (see Eq.~\eqref{eq:phiZnMn}). Results for $\alpha$-\ce{ZnMn2O4} are presented with respect to \acs{VMn1} / \acs{VMn2} sites. The \ce{Mn}-vacant structures before and after relaxation are presented in Figs.~S6~and~S7 in the \ac{SI}.}
\label{tab:VacResults}
\centering
\begin{tabular*}{0.6\textwidth}{c|c|cc|cc}
\cline{1-6}
\multirow{2}{*}{Property} & \acs{phiZnMn} & \multicolumn{2}{c|}{$\alpha$} & \multicolumn{2}{c}{$\lambda$} \\  \cline{3-6}
 & (V~vs~\ce{Zn}/\ce{Zn^{2+}}) & \ce{MnO2} & \ce{ZnMn2O4} & \ce{MnO2} & \ce{ZnMn2O4} \\ \cline{1-6}
\acs{EdVMn0} & 0.762 & 6.9 & -0.1 / 0.2 & 7.9 & 3.0\\
\acs{dErelax} & - & -1.0 & -4.1 / -4.9 & -0.1 & -1.9 \\
\acs{EdVMn0relax} & 0.762 & 7.9 & 4.0 / 5.1 & 8.0 & 4.9\\ \cline{1-6}
\acs{EdVMn-2CBE} & - & 5.4 & -0.1 / 0.2 & 4.4 & 1.7\\ \cline{1-6}
\acs{EdVMn0} & 1.8 & 4.8 & -2.1 / -1.9 & 5.8 & 0.9\\
\acs{EdVMn0} & 1.1 & 6.2 & -0.7 / -0.5 & 7.2 & 2.3\\
\end{tabular*}
\end{table*}

A considerably lower \ac{EdVMn0} value is observed for the vacancy in both \ce{ZnMn2O4} structures. For example, in the $\lambda$ polymorph a reduction of 4.9~eV on the \ac{VMn0} formation energy is observed after the inclusion of \ce{Zn} atoms in the $\lambda$-\ce{MnO2} spinel structure (see Table~\ref{tab:VacResults}). Yet, the calculated \ac{EdVMn0} in $\lambda$-\ce{ZnMn2O4} is still considerably high, which would also restrain the \ce{Mn} dissolution as \ce{Mn^{2+}_{(aq)}} from the structure. However, drastically lower \ac{EdVMn0} values were found for $\alpha$-\ce{ZnMn2O4}, with even negative \ac{EdVMn0} values being established for the \ac{VMn1} formation. The negative \ac{EdVMn0} result indicates the energetic feasibility for \ac{VMn1} formation in $\alpha$-\ce{ZnMn2O4} at an electrode potential of 0.762~V~vs~\ce{Zn}/\ce{Zn^{2+}} (equivalent to 0~V~vs~SHE), demonstrating that the \ce{Mn} atoms become more prone to dissolution as \ce{Mn^{2+}_{(aq)}} once the $\alpha$-\ce{ZnMn2O4} structure is formed. Therefore, the \ac{EdVMn0} results indicate that the \ce{Zn^{2+}} intercalation process occurring during discharge would prompt the transition from an electrochemically stable $\alpha$-\ce{MnO2} structure in the electrode to an unstable $\alpha$-\ce{ZnMn2O4} phase, from which the \ce{Mn} atoms would then dissolve into the electrolyte. The electrochemical instability and associated tendency for \ce{Mn} dissolution from $\alpha$-\ce{ZnMn2O4} under \ac{RAZIB} operation conditions is discussed later in this section.

Now, it is important to understand why the \ac{EdVMn0} is considerably lower for the \ce{ZnMn2O4} structures with respect to their respective \ce{MnO2} polymorphs. In order to answer this question, we have performed \acf{SC} calculations for the structures containing \ac{VMn0} before and after relaxation to quantify the impact of the structural transformations occurring during relaxation on the calculated \ac{EdVMn0}. The \acf{dErelax} is a new parameter established in this work to aid in the \ac{EdVMn0} analysis, capturing the change in the total energy of the vacant system before and after relaxation. \Ac{dErelax} allows for the quantification of the energy reduction in the \ac{VMn0}-containing system due to the structural transformations undergone during relaxation. By subtracting \ac{dErelax} from \ac{EdVMn0}, it is also possible to calculate the \acl{EdVMn0relax} (\acs{EdVMn0relax} $=$ \ac{EdVMn0} $-$ \ac{dErelax}), which is the associated \ac{VMn0} formation energy if no atomic relaxation was performed for the vacant structure. A configuration energy diagram demonstrating the relationship between \ac{EdVMn0}, \acs{EdVMn0relax}, and \ac{dErelax} is presented in Fig.~S5. The \ac{dErelax} and \acs{EdVMn0relax} values calculated for all materials are also shown in Table~\ref{tab:VacResults}, while the \ac{VMn0}-containing structures before and after relaxation are presented in Figs.~S6~and~S7 in the \ac{SI}.

The \acs{EdVMn0relax} results revealed the existence of a common \ac{VMn0} formation energy only dependent on the chemistry of the system, as both \ce{MnO2} structures report a high \acs{EdVMn0relax} value of approximately~8~eV, while the \ce{ZnMn2O4} counterparts have a lower formation energy of about~5~eV. In order to understand the difference between the \acs{EdVMn0relax} results in \ce{MnO2} and \ce{ZnMn2O4} it is necessary to examine the variation in electron availability due to the presence of a \ac{VMn0}. The formation of a \ac{VMn0} will cause a reduction in the number of electrons available in the structure for the atoms to establish their preferential oxidation states. The higher the oxidation state is prior to the vacancy formation, the harder it will be for the atoms to accommodate the electron loss, since more electrons were previously participating in establishing the bonds in the structure. Therefore, the high \ce{Mn} oxidation state in \ce{MnO2} (\ce{Mn}$^{\mathrm{IV}}_{\mathrm{(s)}}$) will make the \ac{VMn0} formation less favourable (higher \acs{EdVMn0relax})  than in \ce{ZnMn2O4} (\ce{Mn}$^{\mathrm{III}}_{\mathrm{(s)}}$), as there will be a lower number of electrons available in the vacant structure for the \ce{Mn} to maintain the considerably high IV oxidation state. In order to verify the proposed relationship between oxidation state and \acs{EdVMn0relax}, the formation of a \ac{VMn0} was also investigated for \ce{MnO} (\ce{Mn}$^{\mathrm{II}}_{\mathrm{(s)}}$) and \ce{Mn2O3} (\ce{Mn}$^{\mathrm{III}}_{\mathrm{(s)}}$). The \acs{EdVMn0relax} results for different \ce{Mn} oxidation states is presented in Fig.~S8, where a clear trend is established between the results for the binary \ce{Mn} oxides (\textit{i.e.}, \ce{MnO}, \ce{Mn2O3} and \ce{MnO2}) and \ce{ZnMn2O4}, showing that the higher the \ce{Mn} oxidation state, the higher the \acs{EdVMn0relax} is, and that \ce{Mn} atoms with the same oxidation state (in this case \ce{Mn}$^{\mathrm{III}}_{\mathrm{(s)}}$) indeed have similar \acs{EdVMn0relax} results. Finally, the same trend between oxidation state and \acs{EdVMn0relax}, as discussed here for \ce{Mn}, can also be expected for the vacancy formation of other elements in other chemical systems, since the rationale behind the trend is independent of the \ce{Mn} oxide chemistry being studied here. For example, the observed trend can be directly applied for screening new stable cathode materials for batteries, as materials with redox-active atoms at higher oxidation state will be less prone to dissolution during cycling.

The \acs{EdVMn0relax} base values calculated for \ce{MnO2} (ca.~8~eV) and \ce{ZnMn2O4} (ca.~5~eV) can then be interpreted as the upper limit for the \ac{VMn0} formation energy in the respective materials, as the chemistry was shown to be the primary parameter governing the vacancy formation energy when relaxation is not considered. Then, it is possible to conclude that the \ce{ZnMn2O4} structures are invariably more prone to \ce{Mn} dissolution than the respective \ce{MnO2} polymorph due to a lower \ce{Mn} oxidation state established after \ce{Zn^{2+}} intercalation. Similar results can be expected for cathode materials in other battery chemistries, with a lower vacancy formation energy being found for the discharged phase (\textit{i.e.}, higher propensity for active material dissolution) than for the charged phase due to the reduction in transition metal oxidation state after ionic intercalation. Once relaxation is performed, the vacancy formation energy universally decreases, but by different magnitudes, which causes materials with the same chemistry to then have noticeably distinct \ac{EdVMn0} depending on their structure (see Table~\ref{tab:VacResults}). The impact of relaxation on the calculated \ac{EdVMn0} can be clearly seen from the results for different polymorphs of \ce{ZnMn2O4}, as a considerable energy reduction due to the structural transformations undergone during relaxation was observed for $\alpha$-\ce{ZnMn2O4}(\ac{VMn}) (calculated $\lvert\ac{dErelax}\rvert$ greater than 4~eV). As previously discussed, the major structural transformation observed after the relaxation of $\alpha$-\ce{ZnMn2O4}(\ac{VMn}) structures is a shift in \ce{Zn} coordination from bent (\ce{ZnO2}) to octahedral (\ce{ZnO6}), with the \ce{Zn} atom substituting the dissolved \ce{Mn} atom on the \ac{VMn} site (see Fig.~\ref{fig:VacancyStructures}b,c).

A question then arises: how can a change in coordination of a single \ce{Zn} atom in $\alpha$-\ce{ZnMn2O4}(\ac{VMn}) be responsible for an energy decrease from cell relaxation (\ac{dErelax}) of more than 4~eV? To answer this question it is necessary  to look into the energetically favourable coordination environments for \ce{Zn} in solid materials. A literature survey reveals that \ce{Zn} atoms are preferably coordinated tetrahedrally (\ce{Zn\textit{X}4}, e.g., Sphalerite / Wurtzite (\ce{ZnS}), Zincite (\ce{ZnO}), Hemimorphite (\ce{Zn4Si2O7(OH)2.H2O}), Willemite (\ce{Zn2SiO4})) or octahedrally (\ce{Zn\textit{X}6}, e.g., Smithsonite (\ce{ZnCO3}), Hopeite (\ce{Zn3(PO4)2})) in naturally occurring inorganic materials~\cite{Barak_Zinc_1993, Neumann_MMJMS_28_1949, Wedepohl_HandGeo_1_1969}. The energetic preference for tetrahedral or octahedral \ce{Zn} coordinations can be ascribed to the complete filling of the degenerate $\text{t}_{2g}$ and $\text{e}_{g}$ states in the \textit{d}-block molecular orbitals for both coordinations by the \ce{Zn^{2+}} $d^{10}$ electrons~\cite{Jean_MOTMC_2005}. In the case of $\lambda$-\ce{ZnMn2O4}(\ac{VMn}), the \ce{Zn} atoms are already at an energetically favourable tetrahedral coordination in the \ac{VMn}-free structure (see Fig.~S2b), which will cause for the material to have a lower \ac{dErelax} since no considerable structural transformations are energetically favourable to occur. A $\lambda$-\ce{ZnMn2O4}(\ac{VMn}) cell with a \ce{Zn} atom octahedrally coordinated in the \ac{VMn} site was also investigated, with negligible change in energy with respect to the tetrahedral \ce{Zn} coordination being found ($<$~50~meV). However, in $\alpha$-\ce{ZnMn2O4}, the \ce{Zn} atoms are in a bent coordination (\ce{ZnO2}), which, according to the literature~\cite{Barak_Zinc_1993, Neumann_MMJMS_28_1949, Wedepohl_HandGeo_1_1969}, is an energetically unfavourable chemical environment for the \ce{Zn} atoms when compared to the octahedral and tetrahedral coordinations. Therefore, with the formation of a \ac{VMn} in $\alpha$-\ce{ZnMn2O4} the \ce{Zn} atom can shift from a bent to an octahedral coordination in the structure, greatly reducing the energy of the system (high $\lvert\ac{dErelax}\rvert$), and consequently granting a considerably low \ac{EdVMn0} for the $\alpha$-\ce{ZnMn2O4}(\ac{VMn}) structures.

\begin{figure}
\centering
\includegraphics[width=0.43\textwidth]{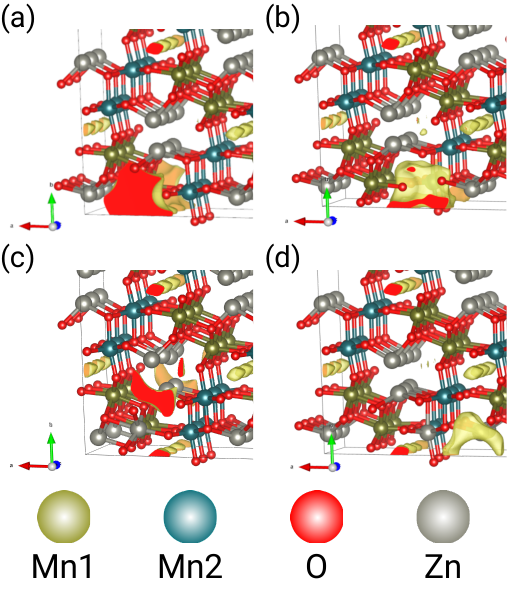}
\caption{(a,c) $\alpha$-\ce{ZnMn2O4}(\ac{VMn1}) and (b,d) $\alpha$-\ce{ZnMn2O4}(\ac{VMn2}) structures (a,b) before and (c,d) after atomic relaxation, featuring the regions of \ac{locpot}~$>$~10~eV in each structure.}
\label{fig:LocalPotential}
\end{figure}

%_%
The \acf{locpot} for the $\alpha$-\ce{ZnMn2O4}(\ac{VMn}) structures before and after relaxation was also calculated in order to understand possible driving factors promoting the change in \ce{Zn} atom coordination. The \ac{locpot} analysis display the regions inside the  \ac{VMn0}-containing supercell where potential energy is concentrated, allowing for the visualization of regions that would be energetically favourable for ions to migrate towards during relaxation. In the \ac{DFT} calculation, all properties are calculated with respect to the electron from the resulting electron density arrangement in the material. For this reason, a positive \ac{locpot} concentration would be attractive for positively charged species (\textit{e.g.}, \ce{Zn^{2+}}) and a negative \ac{locpot} concentration for negatively charged species (\textit{e.g.}, $\mathrm{e^{-}}$). The \ac{locpot} is calculated at each point in the 3D real-space grid; therefore, to facilitate the visualization of cell regions with local potential concentration, it is necessary to determine a minimum cutoff value for the \ac{locpot} points to be displayed. The calculated \ac{locpot} presented here considers only the ionic and Hartree potentials (LVHAR=.TRUE. tag). The resulting volumetric regions for which \ac{locpot}~$>$~10~eV in both $\alpha$-\ce{ZnMn2O4}(\ac{VMn}) structures before and after relaxation are shown in Fig.~\ref{fig:LocalPotential}. A \ac{locpot} concentration around the \ac{VMn} site can be clearly seen in both $\alpha$-\ce{ZnMn2O4}(\ac{VMn}) structures before performing cell relaxation (Fig.~\ref{fig:LocalPotential}a,b), indicating that the empty \ac{VMn} site is attractive for a \ce{Zn^{2+}} cation. A \ce{Zn^{2+}} ion then octahedrally coordinates at the \ac{VMn} site once the relaxation is conducted, with a region of a high \ac{locpot} being established on the site previously occupied by the \ce{Zn^{2+}}, as expected (Fig.~\ref{fig:LocalPotential}c,d). Therefore, the positive \ac{locpot} concentration on the \ac{VMn} site is responsible for attracting the bent coordinated \ce{Zn^{2+}} ion in $\alpha$-\ce{ZnMn2O4}(\ac{VMn}) to the \ac{VMn} site, with the established octahedral coordination causing for the energy of the system to be greatly reduced (high $\lvert\ac{dErelax}\rvert$). A similar \ac{locpot} analysis was also performed for $\alpha$-\ce{MnO2}(\ac{VMn}), $\lambda$-\ce{MnO2}(\ac{VMn}), and $\lambda$-\ce{ZnMn2O4}(\ac{VMn}), with negligible variations in \ac{locpot} being observed before and after relaxation (see Fig.~S9) 

The formation of charged \ac{VMn} defects ($q=-2$) was also investigated in all oxide structures, with the charged defect formation energy calculated from Eq.~\eqref{eq:FormEnergyDefect_final}. The results for the \ac{EdVMn} as function of the \ac{EF} are presented in Fig.~S10. For all materials, except $\alpha$-\ce{ZnMn2O4}(\ac{VMn}), the transition energy level between the neutral and negatively charged \ac{VMn} formation was calculated to be within the band gap for the respective host structures. The neutral to charged \ac{VMn} transition energy level in $\alpha$-\ce{ZnMn2O4} was established for both \ce{Mn}1 and \ce{Mn}2 sites at the \ac{EF} of the host structure (Fig.~S10b), as expected for a material with a metallic character. Lower \ac{EdVMn} values than the one calculated for \ac{VMn0} can be attained for both the $\alpha$- and $\lambda$-\ce{MnO2}, depending upon the position of \acs{dEF}. The lowest \ac{EdVMn} would be at the \ac{CBE} for both $\alpha$-\ce{MnO2}(\ac{VMn}) and $\lambda$-\ce{MnO2}(\ac{VMn}), with the \ac{EdVMn-2CBE} results for all materials also being shown in Table~\ref{tab:VacResults}. It is important to note that even the lowest \ac{EdVMn} results found for the \ce{MnO2} structures are still significantly high (greater than~4~eV), clearly demonstrating how energetically unfavourable the formation of a \ac{VMn} in \ce{MnO2} is. Similar conclusions can also be reached for the formation of a \ac{VMn} in $\lambda$-\ce{ZnMn2O4}, since the lowest calculated \ac{EdVMn}, also at the \ac{CBE}, is still higher than~1.5~eV (see Table~\ref{tab:VacResults}). 
The transition level between neutral and charged \ac{VMn} for $\alpha$-\ce{ZnMn2O4} being established at the \acs{EF} leads to the \ac{EdVMn} value for $\alpha$-\ce{ZnMn2O4}(\ac{VMn}) being independent of the defect charge state considered (\ac{EdVMn0} = \ac{EdVMn-2CBE}, as shown in Table~\ref{tab:VacResults}). In conclusion, the higher propensity for \ce{Mn} dissolution in the $\alpha$-\ce{ZnMn2O4} structure than in the other oxides evaluated is still validated after considering the formation of charged \ac{VMn} defects.

The applied electrode potential continuously varies during battery cycling, which directly impacts the calculated \ac{VMn0} formation energy for each material (see Eq.~\eqref{eq:FormEnergyDefect_final}). Therefore, to determine the energetic feasibility for \ce{Mn} dissolution from each phase during battery operation it is necessary to consider the potential window relevant for \ac{RAZIB} cycling. When \ac{EdVMn0} is equal to 0~eV, the \ce{Mn} vacancy formation is in an energetic equilibrium, with the associated electrode potential being the \ac{phiZnMnrev}. Also, according to Eq.~\eqref{eq:FormEnergyDefect_final}, if the \ac{phiZnMn} is higher than \ac{phiZnMnrev}, the \ac{VMn0} defect formation is energetically viable (\ac{EdVMn0}~$<$~0~eV, \ce{Mn} is predicted to dissolve as \ce{Mn^{2+}_{(aq)}}, oxidation reaction in Eqs.~\eqref{eq:MnO2_vacancy} and \eqref{eq:ZnMn2O4_vacancy}). And, if \ac{phiZnMn} is lower than \ac{phiZnMnrev} the \ac{VMn0} defect is not predicted to form (\ac{EdVMn0}~$>$~0~eV, \ce{Mn} is not predicted to dissolve and remains on the cathode, reduction reaction in Eqs.~\eqref{eq:MnO2_vacancy} and \eqref{eq:ZnMn2O4_vacancy}). The calculated \ac{phiZnMnrev} values for each material is then presented on Table~\ref{tab:phiZnMnrev}. For $\alpha$-\ce{MnO2}, $\lambda$-\ce{MnO2}, and $\lambda$-\ce{ZnMn2O4}, it is predicted that the \ac{VMn0} formation will be energetically unfeasible to occur throughout the \ce{MnO2} cathode \ac{RAZIB} potential cycling window (from 1.1 to 1.8~V~vs~\ce{Zn}/\ce{Zn^{2+}}~\cite{Zhou_JCIS_605_2022, Miliante_JPCC_128_2024}), since the battery would only cycle at potentials lower than the calculated \ac{phiZnMnrev}. However, in the case of $\alpha$-\ce{ZnMn2O4}, a considerably lower \ac{phiZnMnrev} was calculated for both \ce{Mn}1 and \ce{Mn}2 sites (see Table~\ref{tab:phiZnMnrev}). The \ac{phiZnMnrev} around 0.8~V~vs~\ce{Zn}/\ce{Zn^{2+}} calculated for $\alpha$-\ce{ZnMn2O4} indicates that the dissolution of \ce{Mn} atoms as \ce{Mn^{2+}_{(aq)}} will be thermodynamically favourable throughout the potential window for \ac{RAZIB} cycling. The same conclusion of \ac{VMn} formation in $\alpha$-\ce{ZnMn2O4} throughout the \ac{RAZIB} cycling can be reached from the negative \ac{EdVMn0} results at the cycling boundary potentials presented in Table~\ref{tab:VacResults}. Therefore, it is predicted that at discharge, where $\alpha$-\ce{ZnMn2O4} is the primary phase in the electrode, \ce{Mn} vacancies will be energetically favourable to form, consequently increasing the concentration of \ce{Mn^{2+}_{(aq)}} in the electrolyte.

\begin{table}
\caption{\Acf{phiZnMnrev} in~V~vs~\ce{Zn}/\ce{Zn^{2+}} for the \ce{MnO2} and \ce{ZnMn2O4} polymorphs. Results for $\alpha$-\ce{ZnMn2O4} are presented with respect to \acs{VMn1} / \acs{VMn2} sites.}
\label{tab:phiZnMnrev}
\begin{tabular*}{0.48\textwidth}{c|cc|cc}
\cline{1-5}
\multirow{2}{*}{Property}& \multicolumn{2}{c|}{$\alpha$} & \multicolumn{2}{c}{$\lambda$} \\ \cline{2-5}
& \ce{MnO2} & \ce{ZnMn2O4} & \ce{MnO2} & \ce{ZnMn2O4} \\ \cline{1-5}
\acs{phiZnMnrev} & 4.2& 0.7 / 0.8& 4.7& 2.2\\ 
\end{tabular*}
\end{table}

From the analysis of the \ac{EdVMn0} results at relevant \ac{RAZIB} cycling potentials, it is possible to ascribe the proneness for \ce{Mn} dissolution from $\alpha$-\ce{MnO2} cathodes during cycling to the unstable \ce{Zn} coordination established during intercalation. The conclusion for \ce{Mn} dissolution promoted by unstable \ce{Zn} coordination on the structure can be arrived from the analysis of the structural transformations undergone by $\alpha$-\ce{ZnMn2O4}(\ac{VMn}) during relaxation, which have been shown to drastically reduce the \ac{EdVMn0} for the phase (high $\lvert\ac{dErelax}\rvert$) and ultimately contribute for $\alpha$-\ce{ZnMn2O4}(\ac{VMn}) to have negative \ac{EdVMn0} throughout the \ac{RAZIB} cycling window. The \ce{Mn} dissolution mechanism based on unstable \ce{Zn} coordination occurring during \ac{RAZIB} operation can be described as follow. During discharge, \ce{Zn^{2+}} ions intercalate into the electrode and establish themselves with an unstable bent coordination on the $\alpha$-\ce{MnO2} tunnelled structure. The unstable \ce{Zn} coordination on the structure severely reduces the energy for \ce{Mn} dissolution by promoting a substitution reaction, with a \ce{Mn} atom dissolving into the electrolyte as \ce{Mn^{2+}_{(aq)}} while a \ce{Zn} atom establishes an octahedral coordination on the vacant \ce{Mn} site. Despite being energetically favoured, the $\alpha$-\ce{ZnMn2O4}(\ac{VMn}) structure with an octahedrally coordinated \ce{Zn} atom can not be considered structurally stable. The overall structural instability of $\alpha$-\ce{ZnMn2O4}(\ac{VMn}) comes from the presence of numerous \ce{Zn^{2+}} ions with bent coordination within the tunnelled structure, which has been demonstrated to promote \ce{Mn} dissolution in $\alpha$-\ce{ZnMn2O4} cathodes from our calculations. Therefore, further cathode degradation following the initial \ce{Mn} dissolution due to the unstable \ce{Zn} coordination environment captured for $\alpha$-\ce{ZnMn2O4}(\ac{VMn}) is expected to occur. The structural instability of $\alpha$-\ce{ZnMn2O4}(\ac{VMn}) and the subsequent structures formed in the cathode throughout the discharge process are also expected to promote the overall degradation of the cathode material and consequential dissolution of the octahedrally coordinated \ce{Zn} atom into the electrolyte. The expected dissolution of the \ce{Zn} atoms in the cathode back into the electrolyte is supported by the fact that the main \ce{Zn}-containing phases previously experimentally identified in the cathode after battery discharge (\textit{e.g.}, \ce{Zn4SO4(OH)6}$\cdot$5\ce{H2O} (zinc hydroxide sulphate, ZHS) in \acp{RAZIB} with \ce{ZnSO4} electrolyte)\cite{Wu_EES_13_2020, Rubel_JPCC_126_2022, Pan_NE_1_2016} and charge (\textit{e.g.}, \ce{ZnMn3O7}, $\lambda$-\ce{ZnMn2O4})\cite{Liao_ESM_44_2022, Rubel_JPCC_126_2022, Wu_EES_13_2020} are directly formed from the respective dissolved ions present in the electrolyte (\textit{e.g.}, \ce{Zn^{2+}_{(aq)}}, \ce{Mn^{2+}_{(aq)}}, \ce{SO^{2-}_{4(aq)}}, and \ce{OH^{-}_{(aq)}}), with \ce{Zn^{2+}} not being directly identified coordinated to $\alpha$-\ce{MnO2} due to the instability of the system.

It is important to note that the \ce{Mn} dissolution mechanism described here may not be unique to $\alpha$-\ce{MnO2} electrodes for \acp{RAZIB}, since what causes the \ce{Mn} dissolution to become energetically favourable is the presence of an unstable coordination environment, which can potentially occur in different cathodes and battery chemistries. Also, the impact of unstable \ce{Zn} coordination for promoting the \ce{Mn} dissolution in the tunnelled $\alpha$-\ce{MnO2} structure highlights the importance of considering the intercalating ion coordination environment to the host structure, and not only the available space for ionic diffusion, when exploring novel cathode materials for \ac{RAZIB} and other battery chemistries. For example, \ce{MnO2} polymorphs with wider tunnels than $\alpha$-\ce{MnO2}, such as Romanèchite-\ce{MnO2} (2$\times$3 tunnels) and Todorokite-\ce{MnO2} (3×3 tunnels), will also be susceptive to the unstable \ce{Zn} coordination \ce{Mn} dissolution mechanism being proposed here, since a favourable \ce{Zn} coordination environment is also not expected to be established in the structures due to the extensive tunnel space. Therefore, by considering the effect of unstable coordination of intercalating ions on the electrode material stability under operation conditions, a more rational design of cathode materials to support the ongoing energy transition can be expected.

So far, the formation of the \ac{VMn} has only been investigated for the two extreme battery compositions, the completely charged (\ce{MnO2}) and discharged (\ce{ZnMn2O4}) phases. However, the composition, and consequently structure, of the initially $\alpha$-\ce{MnO2} electrode is actively changing during battery cycling, which will then cause for the \ac{EdVMn0} of the electrode to also vary during cycling. Therefore, it is necessary to capture both the phase present in the electrode and the applied potential in order to accurately model the \ac{EdVMn0} of the electrode. Considering a rocking chair battery model~\cite{Deng_ESM_60_2023}, the major phase present in the electrode at high potentials will be the completely charged phase of $\alpha$-\ce{MnO2}. And during discharge, capacity will be gained by having \ce{Zn^{2+}} intercalating into the electrode, with the \ce{Zn} content in $\alpha$-\ce{Zn_{$x$}Mn2O4} increasing until the completely discharged phase $\alpha$-\ce{ZnMn2O4} is established. The $\alpha$-\ce{Zn_{$x$}Mn2O4} phase present on the electrode can then be actively tracked with respect to the capacity being gained on the electrode, with $x$ considered to be equal to 0 at the top of charge (\textit{i.e.}, $\alpha$-\ce{MnO2}) and 1 at the bottom of discharge (\textit{i.e.}, $\alpha$-\ce{ZnMn2O4}). The \ac{EdVMn0} for the electrode was then calculated through a linear interpolation of the \ac{EdVMn0} results for $\alpha$-\ce{MnO2}(\ac{VMn}), $\alpha$-\ce{Zn_{0.5}Mn2O4}(\ac{VMn}), and $\alpha$-\ce{ZnMn2O4}(\ac{VMn1}) with respect to the experimental \ac{phiZnMn} and capacity gained for an $\alpha$-\ce{MnO2} electrode. A partially discharged $\alpha$-\ce{Zn_{0.5}Mn2O4} ($x$~=~0.5) structure was also considered to more accurately capture the \ac{EdVMn0} profile for the electrode during cycling, with 24 out of the 48 atoms present in $\alpha$-\ce{ZnMn2O4} being accounted for during the creation of the $\alpha$-\ce{Zn_{0.5}Mn2O4} supercell. The large number of crystallographically distinct $\alpha$-\ce{Zn_{0.5}Mn2O4} structures that can be generated from the $\alpha$-\ce{ZnMn2O4} template made exploring all possible $\alpha$-\ce{Zn_{0.5}Mn2O4} structures computationally prohibitive. Therefore, random removal of \ce{Zn} atoms from $\alpha$-\ce{ZnMn2O4} was chosen for the creation of unbiased $\alpha$-\ce{Zn_{0.5}Mn2O4} representative structures. In total, three $\alpha$-\ce{Zn_{0.5}Mn2O4} structures were created through random removal of \ce{Zn} atoms, with \ac{VMn} formation in three different \ce{Mn} sites for each $\alpha$-\ce{Zn_{0.5}Mn2O4} structure being investigated to obtain a more accurate calculation of the \ac{EdVMn0} for $\alpha$-\ce{Zn_{0.5}Mn2O4}(\ac{VMn}). The  \ac{EdVMn0} result for $\alpha$-\ce{Zn_{0.5}Mn2O4}(\ac{VMn}) utilized in the calculation of the \ac{EdVMn0} for the electrode was then taken as the average \ac{EdVMn0} result for all nine $\alpha$-\ce{Zn_{0.5}Mn2O4}(\ac{VMn}) systems explored.

\begin{figure}
\centering
\includegraphics[width=0.4\textwidth]{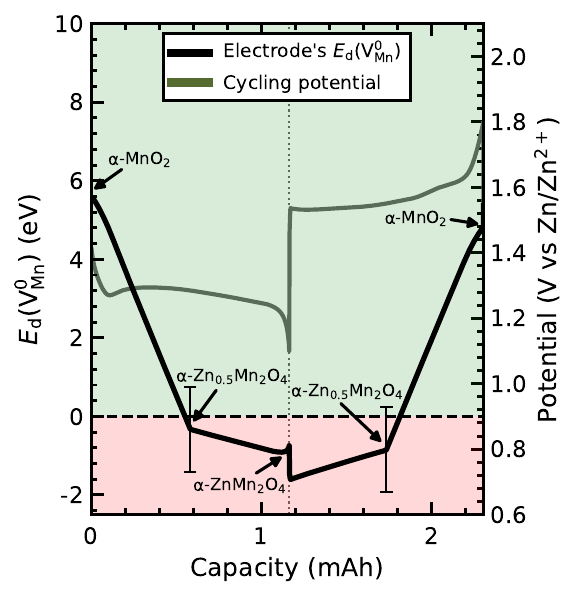}
\caption{The \acf{EdVMn0} for the electrode calculated with respect to the $\alpha$-\ce{MnO2} cathode \acf{GCD} experiment result (initial active material loading equal to 4.5~mg). The experimental \ac{GCD} cycling potential profile used for calculating \ac{EdVMn0} is also shown in the figure on a secondary axis. The green and red regions highlight the \ac{EdVMn0} results for which the \ce{Mn} dissolution is respectively thermodynamically unfavourable (\ac{EdVMn0} $>$ 0~eV) and favourable (\ac{EdVMn0}~$<$~0~eV). The specific capacity shown in the figure is the accumulated specific capacity expended and restored through the discharge and charge cycles. The arrows present in the graph highlight the positions in the \ac{EdVMn0} curve for the electrode where the $\alpha$-\ce{MnO2}, $\alpha$-\ce{Zn_{0.5}Mn2O4}, and $\alpha$-\ce{ZnMn2O4} phases are considered to constitute the electrode. Error bars are added to the results for $\alpha$-\ce{Zn_{0.5}Mn2O4} based on the calculated standard deviation results for the investigated $\alpha$-\ce{Zn_{0.5}Mn2O4}(\ac{VMn}) systems.}
\label{fig:ElectEdVMn}
\end{figure}

The predicted \ac{EdVMn0} for the electrode during battery cycling is shown in Fig.~\ref{fig:ElectEdVMn}, alongside the experimental \acf{GCD} profile for a \ac{RAZIB} cell with an $\alpha$-\ce{MnO2} cathode. As explained previously, multiple reports rely on the \ac{JTE}-assisted charge disproportionation reaction from \ce{ZnMn2O4} to explain the \ce{Mn} dissolution in $\alpha$-\ce{MnO2} RAZIB cathodes as \ce{Mn^{2+}_{(aq)}}~\cite{Lee_SR_4_2014, Tran_SR_11_2021, Siamionau_JPS_523_2022, Cai_CEJ_396_2020, Zhang_NatComms_8_2017, Heo_BS_4_2021, Li_MH_11_2024, Zhang_CPBE_260_2023}. However, our results for the electrode \ac{EdVMn0} show that the \ce{Mn} dissolution is already predicted to be thermodynamically feasible before the battery is fully discharged and the $\alpha$-\ce{ZnMn2O4} phase is completely formed, as the \ac{EdVMn0} results for a partially discharged cell ($x$~$<$~0.5) are already negative. Therefore, there is a slight discrepancy between our results and previous literature that pointed towards the \ac{JTE}-assisted charge disproportionation reaction as the driving mechanism for the \ce{Mn} dissolution. In order to uncover the dissolution dynamics during battery cycling, the \ac{Mn2+conc} on the electrolyte was quantified during cycling of a \ac{RAZIB} cell with $\alpha$-\ce{MnO2} cathode utilizing \textit{operando} \ac{1HNMR}~\cite{Sanders_CM_38_2026}. In the \ac{1HNMR} experiments, the \ce{^{1}H} signal chemical shift and its associated recovery time are directly related to the \ac{Mn2+conc} in the battery electrolyte, with peaks changes to higher chemical shifts indicating higher \ac{Mn2+conc} (see Sec.~\ref{sec:ExpMethods})\cite{Sanders_CM_38_2026}.

\begin{figure}
\centering
\includegraphics[width=0.35\textwidth]{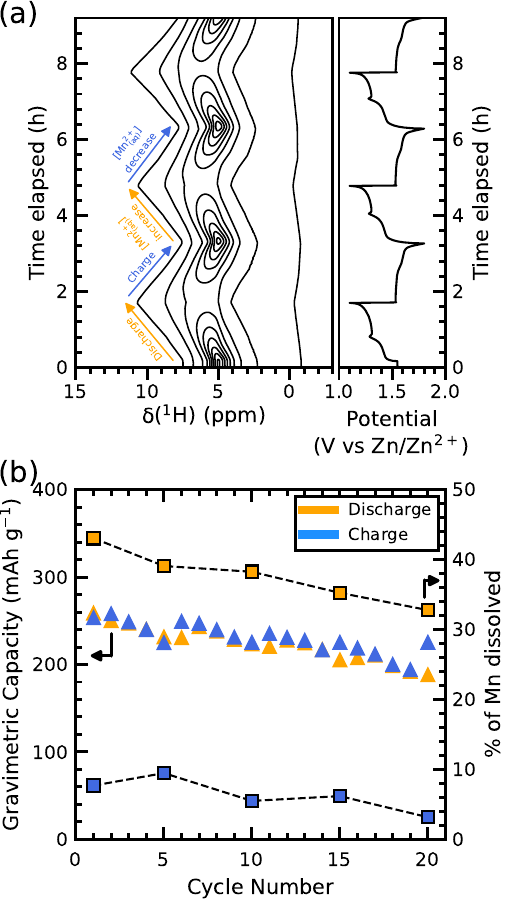}
\caption{(a) Contour plot for chemical shift peak intensity during \textit{operando} \acf{1HNMR} experiment of a \ac{RAZIB} utilizing an $\alpha$-\ce{MnO2} cathode, with its corresponding voltage curve during cycles 2 to 4. (b) Gravimetric capacity and associated \ce{Mn} dissolved percentage with respect to the initial mass in the electrode for the discharged and charged states of the \ac{RAZIB} cell via \textit{in situ} \ac{1HNMR}.}
\label{fig:NMR}
\end{figure}

The \ac{1HNMR} experiments revealed that \ce{Mn} dissolution starts occurring at the beginning of the discharge cycle and continues throughout the entire $\alpha$-\ce{MnO2} cathode discharge, as captured by the increase in chemical shift for the \ce{^{1}H} peak during battery discharge shown in Fig.~\ref{fig:NMR}a. The increase in \ac{Mn2+conc} in the electrolyte at the beginning of the battery discharge, where low concentrations of \ce{Zn^{2+}} ions have intercalated into the structure, indicates the formation of a slightly discharged phase from which \ce{Mn} dissolution is already energetically favourable. Therefore, the \ac{1HNMR} results are in agreement with our theoretical calculations and the proposed unstable \ce{Zn} coordination dissolution mechanism, which predicted that even a partially discharged phase (\textit{e.g.,} $\alpha$-\ce{Zn_{0.5}Mn2O4}) is already electrochemically unstable due to the presence of unfavourably coordinated \ce{Zn^{2+}} ions, causing the \ce{Mn} dissolution to become energetically viable to occur. It should be noted that a similar conclusion of favourable \ce{Mn} dissolution (\textit{i.e.}, negative \ac{EdVMn0} for the electrode) from a partially discharged phase would also be captured by the theoretical results even if the $\alpha$-\ce{Zn_{0.5}Mn2O4} phase had not been considered in our study, but solely the linear interpolation of the $\alpha$-\ce{MnO2}(\ac{VMn}) and $\alpha$-\ce{ZnMn2O4}(\ac{VMn}) results (see Fig.~\ref{fig:ElectEdVMn}). Therefore, the agreement between the theoretical calculations and experimental results is not conditioned by the methodology employed for obtaining the representative $\alpha$-\ce{Zn_{0.5}Mn2O4} structures. The \ac{1HNMR} experiments also show that the \ac{Mn2+conc} in the electrolyte reduces upon charging of the \ac{RAZIB}, as captured by the \ce{^{1}H} peak returns to lower chemical shifts throughout the charge process (see Fig.~\ref{fig:NMR}a). The reduction in \ac{Mn2+conc} during charge is due to the  electrodeposition of \ce{Mn^{2+}_{(aq)}} ions on the electrode as \ce{MnO_{2(s)}}~\cite{Lee_SR_4_2014, Liao_ESM_44_2022, Wu_EES_13_2020}, following the oxidative reaction of the \ce{Mn^{2+}_{(aq)}}/\ce{MnO_{2(s)}} redox couple.

Finally, the gravimetric capacity and percentage of \ce{Mn} dissolved from the electrode during the initial 20 cycles obtained from the \textit{in situ} \ac{1HNMR} experiments are shown in Fig.~\ref{fig:NMR}b. The quantification of the \ce{Mn} dissolution into the electrolyte during battery cycling reveals that more than 40$\mathrm{\%}$ of the \ce{Mn} atoms initially present in the $\alpha$-\ce{MnO2} active material phase dissolve into the electrolyte during the first discharge cycle. It should be noted that the modelled degradation mechanism accounts for the initial \ce{Mn} dissolution from a pristine cathode structure (\textit{i.e.}, without \ce{Mn} vacancies present in the material). Therefore, \ce{Mn} dissolution continues to occur after the initial dissolution promoted by the unstable \ce{Zn} coordination until more than 40$\mathrm{\%}$ of the \ce{Mn} atoms present in the electrode are dissolved. The proposed \ce{Mn} dissolution mechanism via unstable \ce{Zn} coordination is therefore expected to continuously occur during the process of battery discharge, due to the presence of numerous unstable bent-coordinated \ce{Zn^{2+}} ions in the $\alpha$-\ce{Zn_{\textit{x}}Mn2O4}(\ac{VMn}) phase. It is also reasonable to expect that further \ce{Mn} dissolution from the cathode can proceed via other chemical and electrochemical reaction pathways, due to the continuous change in the chemical environment of the cathode material throughout the battery discharge process. However, it is clear from the theoretical results that the unstable coordination of \ce{Zn} will cause the initial dissolution of \ce{Mn} atoms into the electrolyte and set forth the degradation of the \ce{MnO2} cathode. 

The consideration for further electrochemical \ce{Mn} dissolution reactions due to the change in electrode chemical environment promoted by the initial \ce{Mn} dissolution via unstable \ce{Zn} coordination also allows for higher battery capacities to be achieved, surpassing the theoretical limit for the \ce{Zn^{2+}} intercalation mechanism in $\alpha$-\ce{MnO2} (308~mA~h~g$^{-1}$). Experimental support for the additional electrochemical \ce{Mn} dissolution reactions can be found in previous reports that were able to capture battery discharge capacities higher than the theoretical limit associated with \ce{Zn^{2+}} intercalation in $\alpha$-\ce{MnO2} cathodes for \ac{RAZIB}~\cite{Sun_JACS_139_2017, Shi_CSC_14_2021}. On the other hand, a contribution to an increase in specific battery capacity due to \ce{Mn} dissolution can not be achieved if the cathode material undergoes a non-electrochemical dissolution process, such as the \ac{JTE}-assisted charge disproportionation reaction mechanism (Eqs.~\eqref{eq:disprop_reaction}~and~\eqref{eq:disprop_reaction_Zn}), which casts further doubt on the occurrence of the previously proposed mechanism during \ac{RAZIB} cycling with \ce{MnO2} cathodes.

% Exceeding the theoretical capacity should not be possible, unless there is an additional electrochemical reaction occurring alongside the \ce{Zn^{2+}} intercalation that would be able to confer additional capacity. Differently from the \ac{JTE}-assisted charge disproportionation reaction (Eqs.~\eqref{eq:disprop_reaction}~and~\eqref{eq:disprop_reaction_Zn}), the unstable \ce{Zn} coordination \ce{Mn} dissolution mechanism proposed here is an electrochemical reaction (Eqs.~\eqref{eq:MnO2_vacancy}~and~\eqref{eq:ZnMn2O4_vacancy}). Therefore, by considering an electrochemical \ce{Mn} dissolution mechanism, both the \ce{Zn^{2+}} intercalation and \ce{Mn} dissolution would contribute to the battery capacity, allowing for capacities higher than the theoretical limit imposed by \ce{Zn^{2+}} intercalation to be experimentally achieved. 

From the data presented in Fig.~\ref{fig:NMR}b, it is also possible to attest that the majority of \ce{Mn} atoms dissolved into the electrolyte during discharge electrochemically redeposit on the electrode during charge, with less than 10$\mathrm{\%}$ of the \ce{Mn} atoms remaining in the electrolyte after the charging step. The electrochemically deposited phase has been shown to be comprised in its majority of \ce{MnO2}, with \ce{Mn} dissolution from the electrode occurring again in the following discharge cycle~\cite{Lee_SR_4_2014, Liao_ESM_44_2022, Wu_EES_13_2020}. However, inactive phases, such as $\lambda$-\ce{ZnMn2O4}, are also formed in the electrode during the electrodeposition process and do not dissolve into the electrolyte during the following discharge cycles~\cite{Blanc_J_4_2020, Rubel_JPCC_126_2022, Lee_SR_4_2014, Zhang_CPBE_260_2023, Liao_ESM_44_2022}. The formation of a solid phase inactive to the proposed \ce{Mn} dissolution mechanism will also cause a reduction in the obtained gravimetric capacity and \ac{Mn2+conc} after each battery discharge, since a lower amount of \ce{Zn^{2+}} ions will be able to intercalate into the electrode and assist the \ce{Mn} dissolution. The reduction in gravimetric capacity and \ac{Mn2+conc} after successive battery discharges can be clearly seen from the results shown in Fig.~\ref{fig:NMR}b. Overall, the \ce{Mn} dissolution quantification results further showcase the considerable instability of \ac{RAZIB} cells with $\alpha$-\ce{MnO2} cathodes, and highlight the urgency for viable strategies that can address the capacity fade for this material at practical cycling rates for grid-scale energy storage.

\begin{figure*}
\centering
\includegraphics[width=0.8\textwidth]{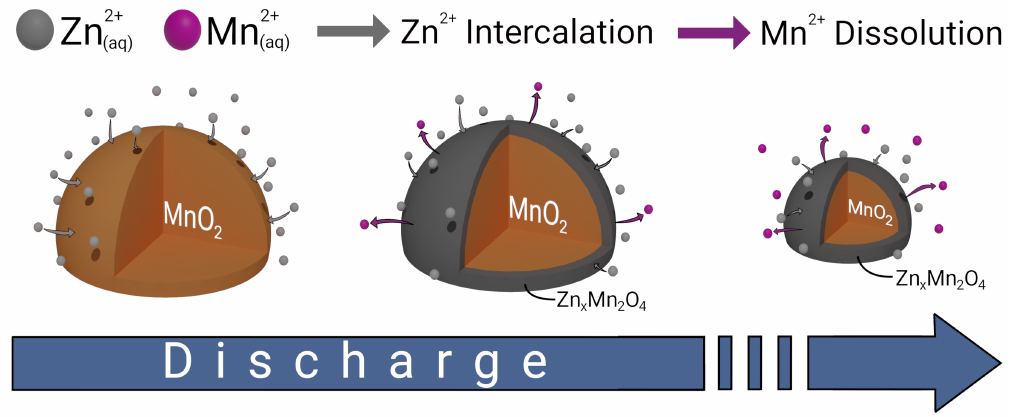}
\caption{Schematic for the \ce{Mn} dissolution process occurring in $\alpha$-\ce{MnO2} particles during \ac{RAZIB} discharge.}
\label{fig:DischargeSchematic}
\end{figure*}

It is possible to conclude from the results presented in this study that the $\alpha$-\ce{MnO2} cathode for \ac{RAZIB} is inherently unstable. First, capacity is gained through \ce{Zn^{2+}} ion intercalation into $\alpha$-\ce{MnO2} during discharge to form a partially discharged $\alpha$-\ce{Zn_{x}Mn2O4} phase. The $\alpha$-\ce{Zn_{x}Mn2O4} phase would then contain \ce{Mn} atoms that have locally undergone a reduction from the IV oxidation state in $\alpha$-\ce{MnO2} to a III oxidation state, as found in $\alpha$-\ce{ZnMn2O4}, due to the intercalation of the \ce{Zn^{2+}} ions in the structure. The intercalated \ce{Zn^{2+}} ions destabilize the electrode structure by establishing an energetically unfavourable bent coordination, which considerably lowers the energy required for \ce{Mn} dissolution and consequently promotes the dissolution of \ce{Mn^{2+}_{(aq)}} into the electrolyte (\textit{i.e.}, the unstable \ce{Zn} coordination \ce{Mn} dissolution mechanism). During charging, the \ce{Mn^{2+}_{(aq)}} ions now present in the electrolyte at higher concentrations will then electrodeposit on the electrode, forming electrochemically active (\textit{e.g.}, \ce{MnO2}) and inactive (\textit{e.g.}, $\lambda$-\ce{ZnMn2O4}) solid phases, with the latter negatively impacting the capacity gain in future cycles and ultimately causing the overall \ac{RAZIB} capacity fade. Therefore, the \ce{Zn^{2+}} intercalation process, which confers the \ce{Zn}/$\alpha$-\ce{MnO2} battery its initial capacity during discharge, is also directly responsible for the battery capacity fade. The experimental \ac{1HNMR} results also showed that the \ce{Mn} dissolution process occurs continuously during discharge, while the theoretical \ac{DFT} results demonstrate that the dissolution can be energetically viable even for partially discharged materials. It is then possible to hypothesize that the \ce{Mn} dissolution process in the $\alpha$-\ce{MnO2} cathode occurs through the formation of an unstable layer of $\alpha$-\ce{Zn_{x}Mn2O4} on the material surface during discharge, from which \ce{Mn} atoms dissolve. As the dissolution occurs, a \ce{MnO2} layer is expected to be established on the surface, with \ce{Zn^{2+}} ions intercalating into the host structure and once again forming the unstable $\alpha$-\ce{Zn_{x}Mn2O4} phase prone to \ce{Mn} dissolution. A schematic for the proposed continuous \ce{Mn} dissolution process occurring during \ac{RAZIB} discharge in a representative $\alpha$-\ce{MnO2} particle is presented in Fig.~\ref{fig:DischargeSchematic}. Further characterization of the formation of the partially reduced $\alpha$-\ce{Zn_{x}Mn2O4} surface phase during cycling and investigation of the attained oxidation states for the \ce{Mn} atoms in the electrode throughout the dissolution process are still necessary to gain additional insight into the \ce{Mn} dissolution and support the development of stable \ce{MnO2} cathodes for \acp{RAZIB}.

It is then necessary to reduce the destabilizing effect that the \ce{Zn} atoms have on the host structure in order to address the capacity fade issue found in $\alpha$-\ce{MnO2} cathodes for \ac{RAZIB}. Based on the theoretical \ac{EdVMn} results for the \ce{ZnMn2O4} systems and the uncovered mechanism for \ce{Mn} dissolution during cycling discussed in the study, two potential strategies are proposed: the inclusion of foreign elements in the tunnelled structure and the utilization of alloying/doping elements on the octahedral \ce{Mn} site. Our results show that the unstable bent coordination established by the \ce{Zn} atom after intercalation into the $\alpha$-\ce{MnO2} tunnel directly contributes to the low \ac{EdVMn} calculated for $\alpha$-\ce{ZnMn2O4}. Therefore, the inclusion of a foreign atom into the tunnelled structure could potentially allow for a more stable coordination (\textit{i.e.}, tetrahedral or octahedral) to be established for the intercalating \ce{Zn^{2+}} ion, consequently increasing the associated \ac{EdVMn}. For example, chemical precursors containing light atoms, such as \ce{Li} and \ce{Be}, can be potentially added during synthesis of $\alpha$-\ce{MnO2} to support the stable intercalation of \ce{Zn^{2+}} ions into the tunnelled structure, only adding a marginal increase in the molecular weight of the active material. It should be noted that the inclusion of a foreign element could potentially impact the intercalation/deintercalation process in structures with smaller tunnels (\textit{e.g.},~$\alpha$-\ce{MnO2} - 2$\times$2 tunnels) by reducing the available space for ionic diffusion inside the material. On the other hand, the capacity fade mitigation strategy of foreign element inclusion can be potentially more successful in \ce{MnO2} polymorphs with wider tunnels (\textit{e.g.}, Romanèchite-\ce{MnO2} - 2$\times$3 tunnels, Todorokite-\ce{MnO2} - 3$\times$3 tunnels), since a lower negative impact of foreign atom inclusion on the reversibility of the \ce{Zn^{2+}} intercalation process would be expected. The chemical environment of the cathode material can also be modified with the incorporation of alloying/doping elements in place of \ce{Mn} in the \ce{MnO2} structure, contributing to an increase in the \ac{EdVMn} after \ce{Zn^{2+}} ions are intercalated. The alloying/doping atoms can potentially constrain the possibility of the \ce{Zn} atoms to octahedrally coordinate at the \ac{VMn} site after the vacancy is formed, which should considerably increase the calculated \ac{EdVMn} in accordance with the \ce{ZnMn2O4} phase results. Even though the \ac{VMn} formation will be invariably more energetically favourable in \ce{Zn}-containing \ce{MnO2} phases than in pristine \ce{MnO2} (see discussion for Table~\ref{tab:VacResults} results), considerably higher \ac{EdVMn} (\textit{i.e.}, lower \ce{Mn} dissolution feasibility) can still be achieved if \ce{Zn^{2+}} is intercalated in a more favourable environment, as can be seen from the results of $\lambda$-\ce{ZnMn2O4}. For example, oxides of \ce{Ti^{IV}} and \ce{W^{VI}} are electrochemically stable under \ac{RAZIB} operation conditions~\cite{Miliante_JPCC_128_2024}, positioning them as ideal ions to be utilized in alloying/doping reactions to replace \ce{Mn} atoms in \ce{MnO2} cathodes. The theoretical methodology developed in this work can be directly utilized by researchers to study the viability and optimal parameters for atomic-level strategies aiming to reduce the \ce{Mn} dissolution during battery discharge, helping to facilitate the experimental implementation of practical cathode modification approaches. The proposed capacity fade mitigation strategies are then an open research avenue for future experimental and theoretical investigations aiming to reduce the impact of \ce{MnO2} cathode instability for the development of commercially viable \acp{RAZIB}.

\section*{Conclusions}

Motivated by the previous uncertainty about the origin of the \ce{Mn} dissolution during cycling of \acp{RAZIB} with \ce{MnO2} cathodes, the formation of \acf{VMn} sites in completely charged ($\alpha$-\ce{MnO2} and $\lambda$-\ce{MnO2}) and discharged ($\alpha$-\ce{ZnMn2O4} and $\lambda$-\ce{ZnMn2O4}) \ce{Mn} oxide phases  was thoroughly investigated. The \ce{Mn} dissolution was determined to be energetically favourable to occur on the discharged phase $\alpha$-\ce{ZnMn2O4} throughout the \ac{RAZIB} potential window utilized for $\alpha$-\ce{MnO2} electrode cycling. The \ce{Mn} dissolution mechanism was concluded to result from the unstable bent coordination from the intercalating \ce{Zn^{2+}} ions during discharge, prompting a substitution reaction to be energetically favourable, where a \ce{Mn} atom dissolves into the electrolyte as \ce{Mn^{2+}_{(aq)}} and a \ce{Zn} atom octahedrally coordinates to the structure on the \ac{VMn} site. The \ac{VMn} formation energy for an $\alpha$-\ce{MnO2} electrode during \ac{RAZIB} cycling was then modelled, with the \ce{Mn} dissolution shown to be energetically viable to occur even with low concentration of \ce{Zn^{2+}} ions intercalated into the electrode ($x$~$<$~0.5 in $\alpha$-\ce{Zn_{x}Mn2O4}). The theoretical results were then corroborated by \textit{operando} \ac{1HNMR} experiments, which captured the continuous dissolution of \ce{Mn} as \ce{Mn^{2+}_{(aq)}} during discharge of a \ac{RAZIB} cell with $\alpha$-\ce{MnO2} cathode. 

The unstable \ce{Zn} coordination \ce{Mn} dissolution mechanism proposed here for $\alpha$-\ce{MnO2} cathodes in \acp{RAZIB} is also applicable for other cathodes materials and battery chemistries, since the unstable coordination of any intercalating ion into a cathode structure could promote the dissolution of the active material. Therefore, consideration for the intercalating ion coordination environment is shown here to be crucial during the investigation of cell degradation and proposal of novel cathode materials for ionic intercalation battery technologies. The discovery of the unstable \ce{Zn} coordination \ce{Mn} dissolution mechanism directly aid the development of novel strategies for capacity fade mitigation in \ac{RAZIB} and other battery chemistries, with two strategies concerning foreign atom inclusion on the structure and utilization of alloying/doping metals being proposed here. Overall, by uncovering origin of the \ce{Mn} dissolution mechanism in $\alpha$-\ce{MnO2} cathodes during \ac{RAZIB} cycling, a more comprehensive understanding of battery cathodes operation and degradation was achieved, directly supporting the design of stable battery technologies with prolonged operation lifetime.

\section*{Author contributions}

\begin{itemize}
\item \textbf{Caio Miranda Miliante}: Conceptualization, Methodology (Theoretical), Validation, Formal analysis, Investigation (Theoretical), Data curation, Writing - Original Draft, Writing - Review \& Editing, Visualization, and Project administration.
\item \textbf{Kevin J. Sanders}: Methodology (Experimental), Investigation (Experimental) and Writing - Review \& Editing.
\item \textbf{Liam J. McGoldrick}: Investigation (Experimental) and Writing - Review \& Editing.
\item \textbf{Nicola Seriani}: Writing - Review \& Editing.
\item \textbf{Brian D. Adams}: Writing - Review \& Editing, and Funding acquisition.
\item \textbf{Gillian R. Goward}: Resources, Writing - Review \& Editing, Supervision, and Funding acquisition.
\item \textbf{Drew Higgins}: Resources, Writing - Review \& Editing, Supervision, and Funding acquisition.
\item \textbf{Oleg Rubel}: Conceptualization, Methodology, Resources, Writing - Review \& Editing, Supervision, Funding acquisition.
\end{itemize}

\section*{Conflicts of interest}

B.D.A. is co-founder and partial owner of Salient Energy Inc., which develops and commercializes rechargeable aqueous \ce{Zn}-ion battery technology.

\section*{Data availability}

The data supporting this article is presented in the \acf{SI}. The \ac{SI} includes data for: \ac{SEM} of $\alpha$-\ce{MnO2} particles; relaxed crystal structures of $\lambda$-\ce{MnO2} and $\lambda$-\ce{ZnMn2O4}; \ac{VMn} site positions in and relaxed structures of $\alpha$-\ce{MnO2}(\ac{VMn}), $\lambda$-\ce{MnO2}(\ac{VMn}), and $\lambda$-\ce{ZnMn2O4}(\ac{VMn}); schematic of the relationship between \ac{EdVMn0}, \ac{dErelax}, and \ac{EdVMn0relax}; structures of $\alpha$-\ce{MnO2}(\ac{VMn}), $\alpha$-\ce{ZnMn2O4}(\ac{VMn1}), $\alpha$-\ce{ZnMn2O4}(\ac{VMn2}), $\lambda$-\ce{MnO2}(\ac{VMn}), and $\lambda$-\ce{ZnMn2O4}(\ac{VMn}) before and after atomic relaxation; \ac{EdVMn0relax} results for \ce{Mn} atoms in different oxidation states with respect to the chemical potential of \ce{Mn^{2+}_{(aq)}} in the \ac{RAZIB} aqueous electrolyte; \ac{locpot} results for $\alpha$-\ce{MnO2}(\ac{VMn}), $\lambda$-\ce{MnO2}(\ac{VMn}) and $\lambda$-\ce{ZnMn2O4}(\ac{VMn})  structures before and after atomic relaxation; \acf{EdVMn} as a function of the \ac{EF} for $\alpha$-\ce{MnO2}(\ac{VMn}), $\alpha$-\ce{ZnMn2O4}(\ac{VMn1}), $\alpha$-\ce{ZnMn2O4}(\ac{VMn2}), $\lambda$-\ce{MnO2}(\ac{VMn}), and $\lambda$-\ce{ZnMn2O4}(\ac{VMn})

\section*{Acknowledgements}

The authors gratefully acknowledge the financial support from  \href{https://salientenergyinc.com/}{Salient Energy Inc.} and the \href{https://www.nserc-crsng.gc.ca/index_eng.asp}{Natural Sciences and Engineering Research Council of Canada (NSERC)} Alliance Program. The \ac{SEM} and \ac{NMR} experiments were respectively conducted at the \href{https://ccem.mcmaster.ca/}{Canadian Centre for Electron Microscopy (CCEM)} and the \href{https://research.science.mcmaster.ca/rcis-and-cores/services/nuclear-magnetic-resonance-facility-nmr/}{Nuclear Magnetic Resonance (NMR) Facility} at McMaster University, Canada. C.M.M. acknowledges the \href{https://www.alliancecan.ca/en}{Digital Research Alliance of Canada} and \href{https://www.computeontario.ca/}{Compute Ontario} for the computing resources utilized in this research. 

%%%END OF MAIN TEXT%%%


\balance


%%%REFERENCES%%%
\bibliography{bibliography} %You need to replace "rsc" on this line with the name of your .bib file
\bibliographystyle{rsc} %the RSC's .bst file
\end{document}

\newcommand{\titlename}{Unveiling the origin of the capacity fade in \ce{MnO2} zinc-ion battery cathodes through an analysis of the \ce{Mn} vacancy formation}
\newcommand{\smtitle}{Supplementary Information: \titlename}

\title{\smtitle}

\author{Caio Miranda Miliante}
\email{miliantc@mcmaster.ca}
\affiliation{Department of Materials Science and Engineering, McMaster University, 1280 Main Street West, Hamilton, Ontario L8S 4L8, Canada}

\author{Kevin J. Sanders}
\author{Liam J. McGoldrick}
\affiliation{Department of Chemistry $\&$ Chemical Biology, McMaster University, 1280 Main Street West, Hamilton, Ontario L8S 4L8, Canada}

\author{Nicola Seriani}
\affiliation{Condensed Matter and Statistical Physics Section, The Abdus Salam ICTP, Strada Costiera 11, 34151 Trieste, Italy}

\author{Brian D. Adams}
\affiliation{Salient Energy Inc., 21 McCurdy Avenue, Dartmouth, Nova Scotia B3B 1C4, Canada}

\author{Gillian R. Goward}
\affiliation{Department of Chemistry $\&$ Chemical Biology, McMaster University, 1280 Main Street West, Hamilton, Ontario L8S 4L8, Canada}

\author{Drew Higgins}
\affiliation{Department of Chemical Engineering, McMaster University, 1280 Main Street West, Hamilton, Ontario L8S 4L8, Canada}

\author{Oleg Rubel}
\affiliation{Department of Materials Science and Engineering, McMaster University, 1280 Main Street West, Hamilton, Ontario L8S 4L8, Canada}

\date{\today}% It is always \today, today,

% \clearpage
%\makeatletter

\renewcommand\thesection{S\arabic{section}}
\renewcommand\thefigure{S\arabic{figure}}
\renewcommand\thetable{S\arabic{table}}
\renewcommand\theequation{S\arabic{equation}}
\renewcommand\thepage{S\arabic{page}}
%\renewcommand\thetitle{Supplementary information for publication}
%\makeatother
\setcounter{section}{0}
\setcounter{figure}{0}
\setcounter{table}{0}
\setcounter{equation}{0}
\setcounter{page}{1}
%\appendix

\maketitle

\vspace*{\fill}
\setlength\extrarowheight{1pt}
\setlength{\tabcolsep}{6pt}
\begin{table}[t]
\caption{Comparison between the experimental and the relaxed supercell crystallographic parameters for $\alpha$-\ce{MnO2}, $\lambda$-\ce{MnO2}, and $\lambda$-\ce{ZnMn2O4}. The parameters reported for the supercells are calculated by dividing the relaxed supercell parameters by the respective number of cell replications utilized when creating the supercell. All materials are from tetragonal or cubic space groups, thus having all lattice angles ($\alpha$, $\beta$, and $\gamma$) equal to 90$\degree$.}
\begin{tabular}{c|c|c|ccc}
\hline
Material & Space Group & Phase & a & b & c \\ \hline
\multirow{2}{*}{$\alpha$-\ce{MnO2}} & \multirow{2}{*}{I4/m} & Supercell (2x2x3) & 9.640 & 9.640 & 2.888 \\ \cline{3-6}
 & & Exp. \cite{Kijima_JSSC_177_2004} & 9.814 & 9.814 & 2.850 \\ \hline
\multirow{2}{*}{$\lambda$-\ce{MnO2}} & \multirow{2}{*}{Fd$\overline{3}$m} & Supercell (2x2x2) & 8.008 & 8.008 & 8.008 \\ \cline{3-6}
 & & Exp. \cite{Hunter_JSSC_39_1981}  & 8.03  & 8.03  & 8.03  \\ \hline
\multirow{2}{*}{$\lambda$-\ce{ZnMn2O4}} & \multirow{2}{*}{I4$_{1}$/amd} & Supercell (2x2x1) & 5.737 & 5.737 & 9.139 \\ \cline{3-6}
 & & Exp. \cite{Menaka_BMS_32_2009}   & 5.709 & 5.709 & 9.238 \\ \hline
\end{tabular}
\label{tbl:CellParam}
\end{table}
\vspace*{\fill}

\vspace*{\fill}
\begin{figure}
\includegraphics[width=0.7\textwidth]{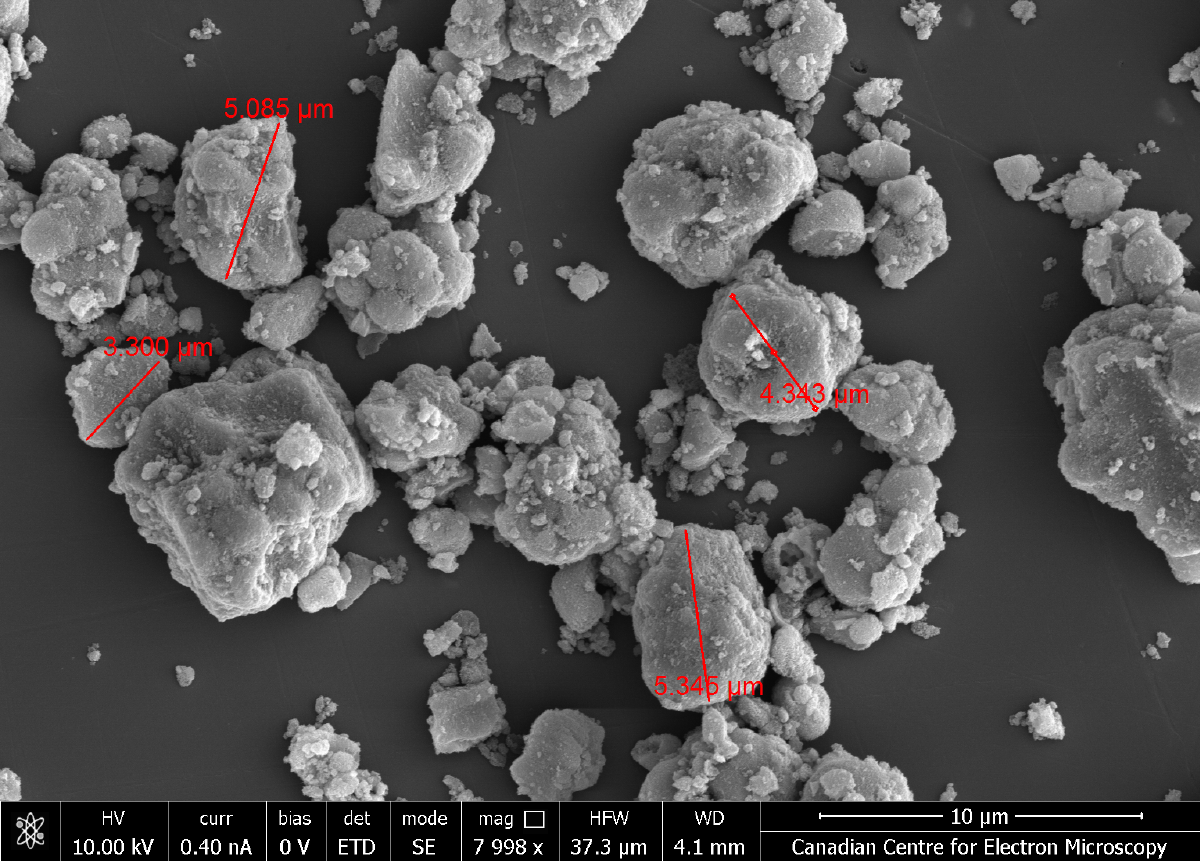}
\caption{\Acf{SEM} image of $\alpha$-\ce{MnO2} active material utilized in experimental testings.}
\label{fig:SEM}
\end{figure}
\vspace*{\fill}

\vspace*{\fill}
\begin{figure}[t]
\includegraphics[width=0.85\textwidth]{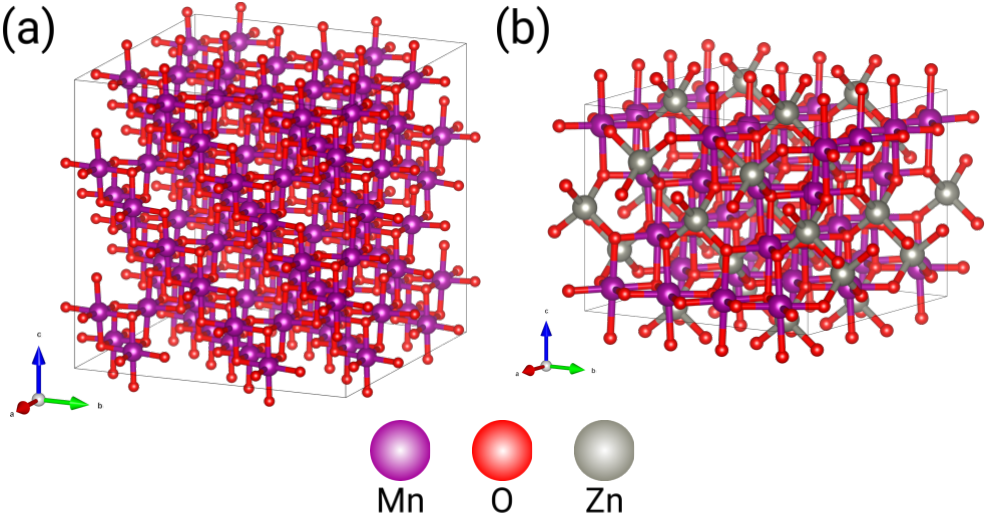}
\caption{Relaxed crystal structures of bulk (a) $\lambda$-\ce{MnO2} and (b) $\lambda$-\ce{ZnMn2O4}.}
\label{fig:Lambda-InitialStructures}
\end{figure}
\vspace*{\fill}

\vspace*{\fill}
\begin{figure}
\includegraphics[width=0.85\textwidth]{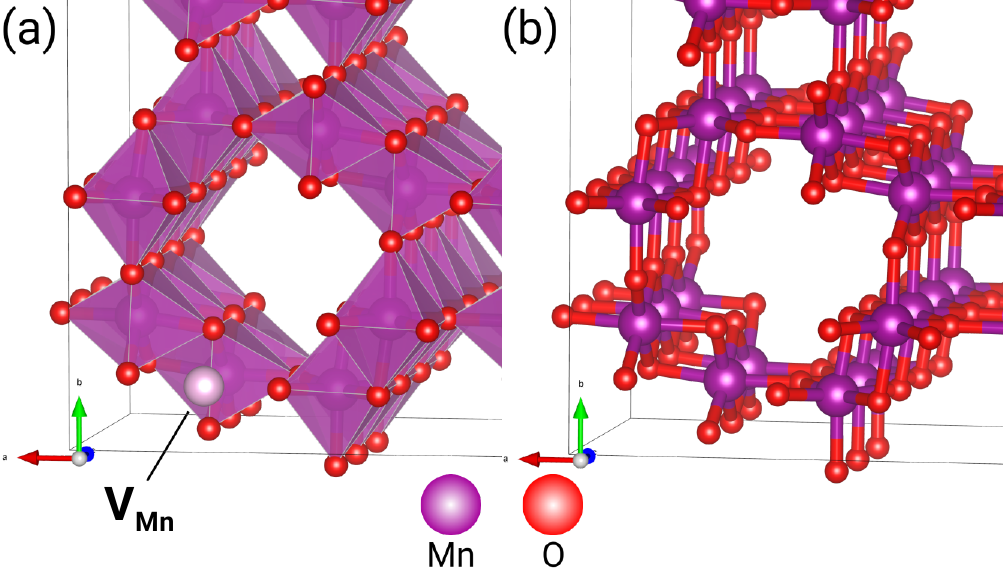}
\caption{(a) \acs{VMn} site in the $\alpha$-\ce{MnO2} crystal structure indicated by lighter shaded \ce{Mn} atom. (b) Relaxed structure of $\alpha$-\ce{MnO2}(\acs{VMn}) for a neutral charge defect. Only a zoomed in region of the  simulated $\alpha$-\ce{MnO2}(\acs{VMn}) cells is shown to facilitate the visual analysis of the \ac{VMn} positioning.}
\label{fig:AlphaMnO2-VacancyStructure}
\end{figure}
\vspace*{\fill}

\vspace*{\fill}
\begin{figure}
\includegraphics[width=0.85\textwidth]{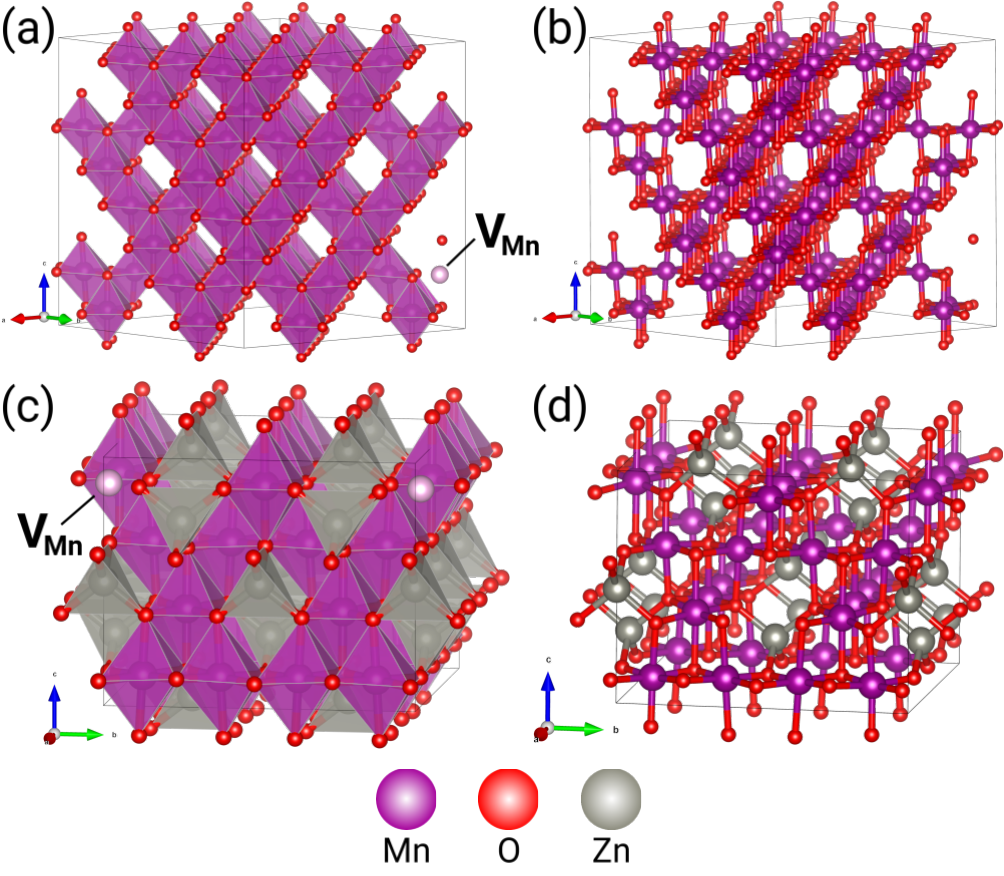}
\caption{\acs{VMn} sites in the (a) $\lambda$-\ce{MnO2} and (c) $\lambda$-\ce{ZnMn2O4} crystal structures indicated by lighter shaded \ce{Mn} atoms. Relaxed structures of (b) $\lambda$-\ce{MnO2}(\acs{VMn}) and (d) $\lambda$-\ce{ZnMn2O4}(\acs{VMn})  for a neutral charge defect.}
\label{fig:Lambda-VacancyStructures}
\end{figure}
\vspace*{\fill}

\vspace*{\fill}
\begin{figure}
\includegraphics[width=0.7\textwidth]{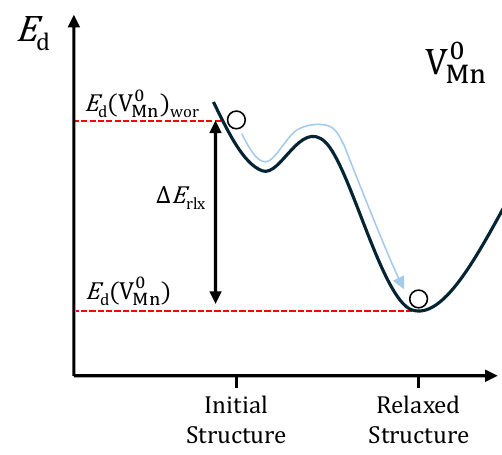}
\caption{Configuration energy diagram depicting the relationship between the \acf{EdVMn0}, the \acf{dErelax}, and the \acf{EdVMn0relax} with respect to the energy profile of a structure containing a \acf{VMn0}.}
\label{fig:NeutDefSchem}
\end{figure}
\vspace*{\fill}

\vspace*{\fill}
\begin{figure}
\includegraphics[width=0.75\textwidth]{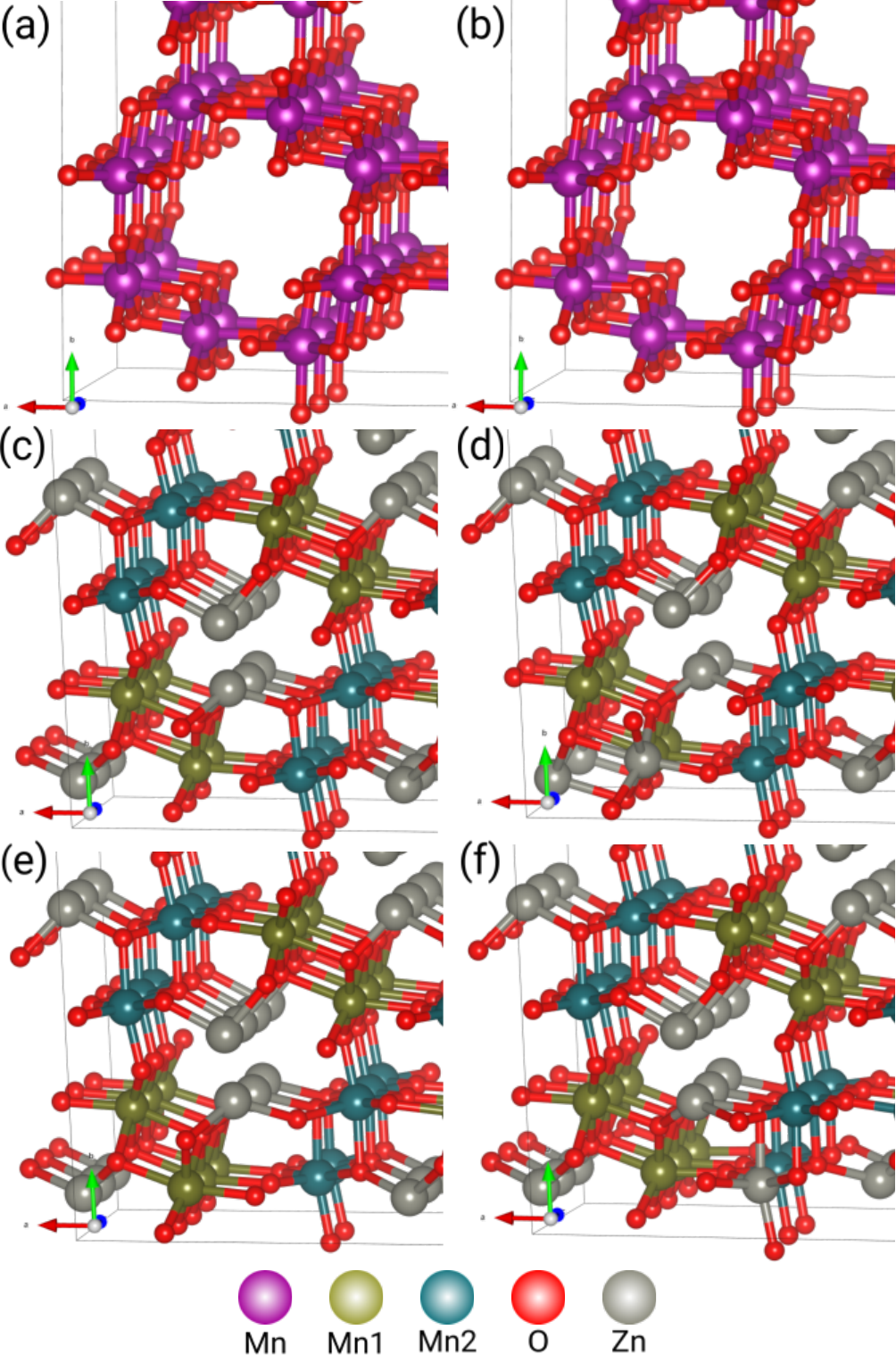}
\caption{Structures of (a,b) $\alpha$-\ce{MnO2}(\acs{VMn}), (c,d) $\alpha$-\ce{ZnMn2O4}(\acs{VMn1}) and (e,f) $\alpha$-\ce{ZnMn2O4}(\acs{VMn2}) before and after performing atomic relaxation.}
\label{fig:Alpha-SC_Structures}
\end{figure}
\vspace*{\fill}

\vspace*{\fill}
\begin{figure}
\includegraphics[width=0.85\textwidth]{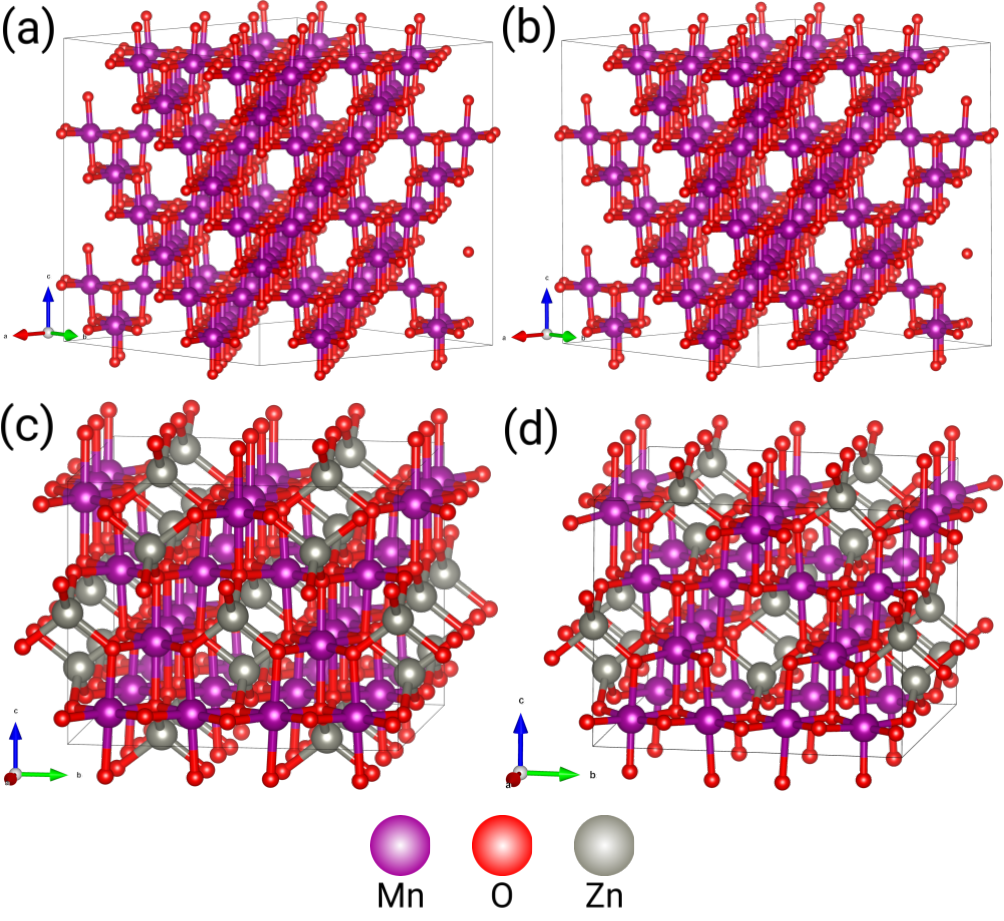}
\caption{Structures of (a,b) $\lambda$-\ce{MnO2}(\acs{VMn}) and (c,d) $\lambda$-\ce{ZnMn2O4}(\acs{VMn}) before and after performing atomic relaxation.}
\label{fig:Lambda-SC_Structures}
\end{figure}
\vspace*{\fill}

\vspace*{\fill}
\begin{figure}
\includegraphics[width=0.7\textwidth]{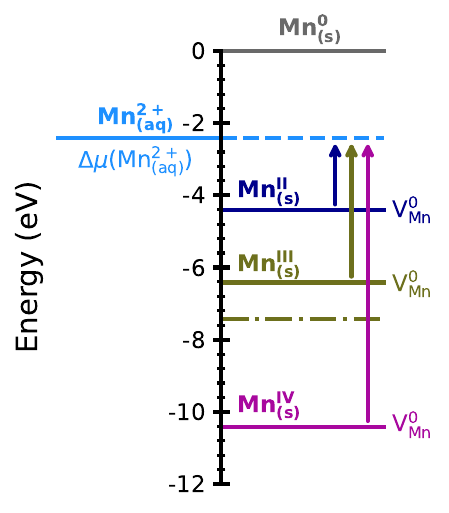}
\caption{Diagram relating the \ac{EdVMn0relax} calculated for \ce{MnO} (\ce{Mn}$^{\mathrm{II}}_{\mathrm{(s)}}$), \ce{Mn2O3} (\ce{Mn}$^{\mathrm{III}}_{\mathrm{(s)}}$) and \ce{MnO2} (\ce{Mn}$^{\mathrm{IV}}_{\mathrm{(s)}}$) to the chemical potential of \ce{Mn^{2+}_{(aq)}} in the \ac{RAZIB} aqueous electrolyte. The dash-dotted line indicates the \ac{EdVMn0relax} result for \ce{ZnMn2O4} (\ce{Mn}$^{\mathrm{III}}_{\mathrm{(s)}}$).}
\label{fig:SchemeEdVMn0relax}
\end{figure}
\vspace*{\fill}

\vspace*{\fill}
\begin{figure}
\includegraphics[width=0.85\textwidth]{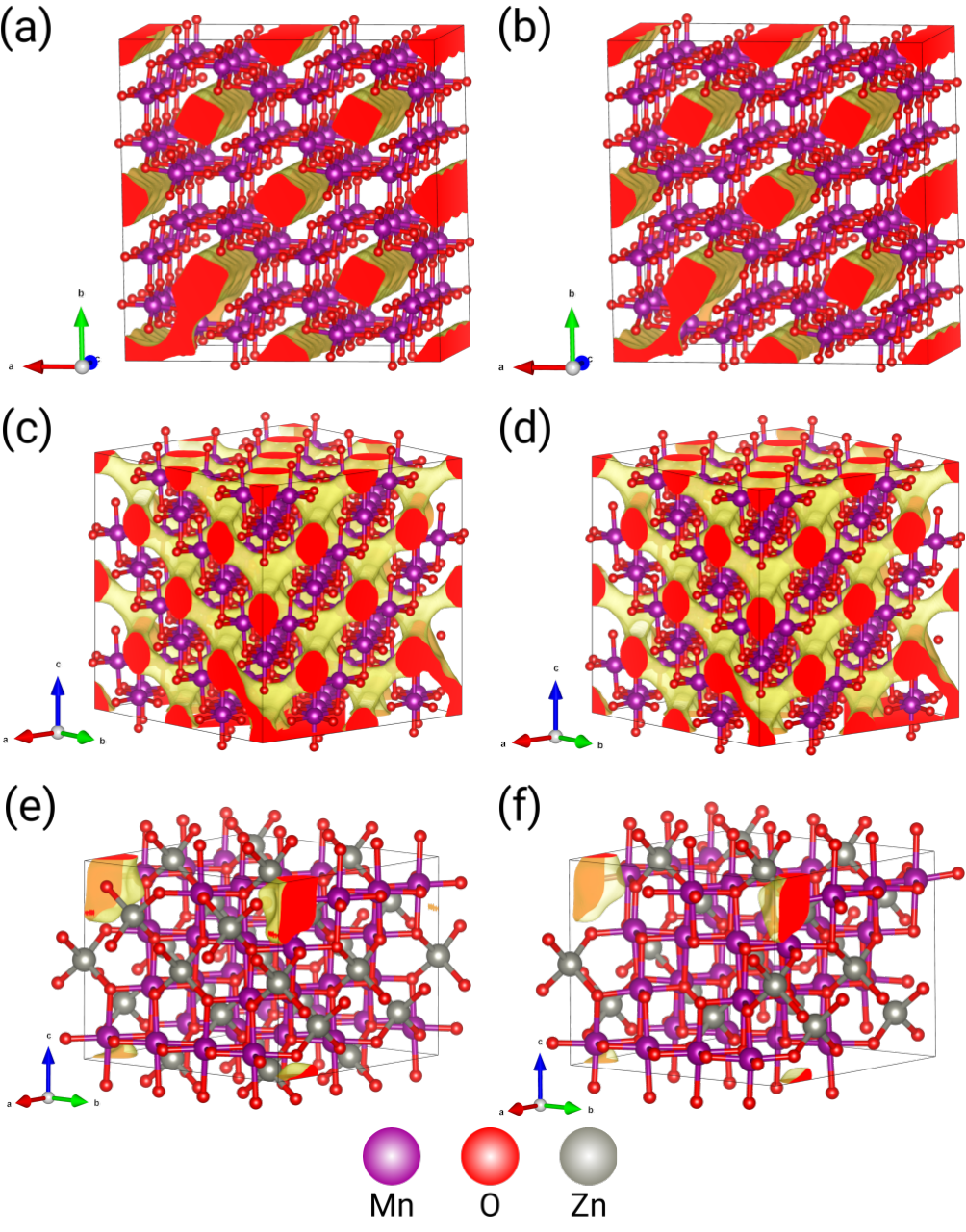}
\caption{Structures of (a,b) $\alpha$-\ce{MnO2}(\ac{VMn}), (c,d) $\lambda$-\ce{MnO2}(\acs{VMn}), and (e,f) $\lambda$-\ce{ZnMn2O4}(\acs{VMn})  before and after atomic relaxation, featuring the regions of \ac{locpot} $>$~10~eV in each structure.}
\label{fig:MnO2-LOCPOT}
\end{figure}
\vspace*{\fill}

\vspace*{\fill}
\begin{figure}
\includegraphics[width=0.85\textwidth]{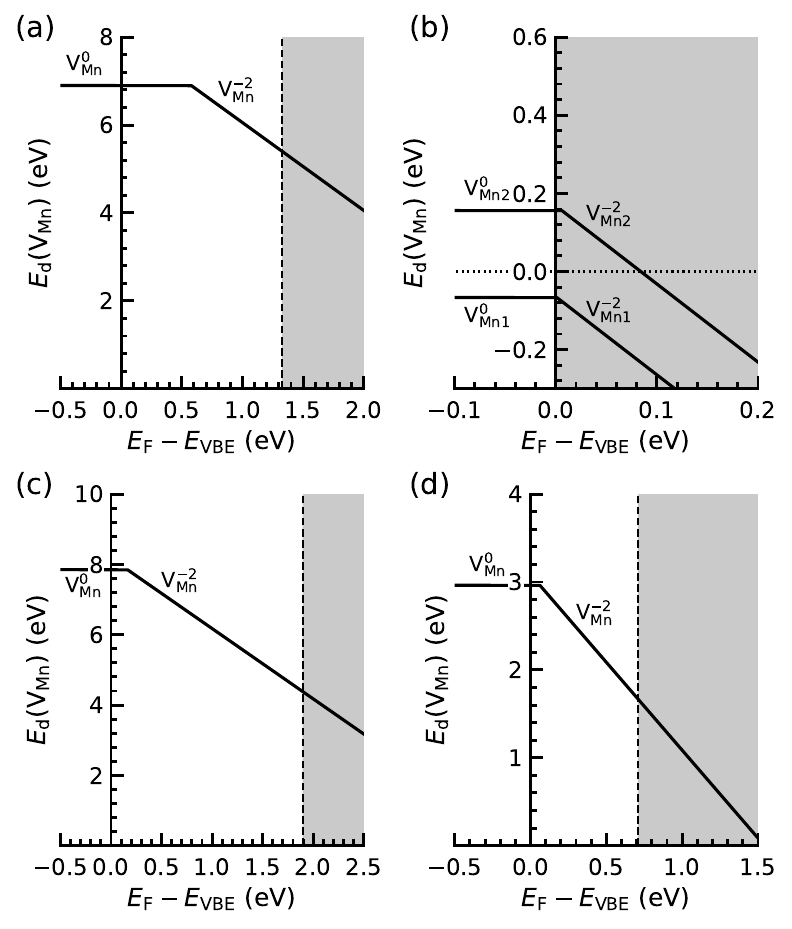}
\caption{\acf{EdVMn} as a function of the \acf{EF} for the (a) $\alpha$-\ce{MnO2}(\acs{VMn}), (b) $\alpha$-\ce{ZnMn2O4}(\acs{VMn}), (c) $\lambda$-\ce{MnO2}(\acs{VMn}), and (d) $\lambda$-\ce{ZnMn2O4}(\acs{VMn}) structures. The vertical dashed lines mark the location of the \acf{CBE}, with the grey zones highlighting conduction band regions.}
\label{fig:TransitionLevel}
\end{figure}

\clearpage
\bibliography{bibliography}